
%
\magnification 1200
\hsize 15.4truecm
\vsize21.8truecm


%
%
%
%

\expandafter\chardef\csname pre amssym.def at\endcsname=\the\catcode`\@
\catcode`\@=11
\def\setboxz@h{\setbox\z@\hbox}
\def\wdz@{\wd\z@}


\def\undefine#1{\let#1\undefined}
\def\newsymbol#1#2#3#4#5{\let\next@\relax
 \ifnum#2=\@ne\let\next@\msafam@\else
 \ifnum#2=\tw@\let\next@\msbfam@\fi\fi
 \mathchardef#1="#3\next@#4#5}
\def\mathhexbox@#1#2#3{\relax
 \ifmmode\mathpalette{}{\m@th\mathchar"#1#2#3}%
 \else\leavevmode\hbox{$\m@th\mathchar"#1#2#3$}\fi}
\def\hexnumber@#1{\ifcase#1 0\or 1\or 2\or 3\or 4\or 5\or 6\or 7\or 8\or
 9\or A\or B\or C\or D\or E\or F\fi}

\font\tenmsa=msam10
\font\sevenmsa=msam7
\font\fivemsa=msam5
\newfam\msafam
\textfont\msafam=\tenmsa
\scriptfont\msafam=\sevenmsa
\scriptscriptfont\msafam=\fivemsa
\edef\msafam@{\hexnumber@\msafam}
\mathchardef\dabar@"0\msafam@39
\def\dashrightarrow{\mathrel{\dabar@\dabar@\mathchar"0\msafam@4B}}
\def\dashleftarrow{\mathrel{\mathchar"0\msafam@4C\dabar@\dabar@}}

\def\ulcorner{\delimiter"4\msafam@70\msafam@70 }
\def\urcorner{\delimiter"5\msafam@71\msafam@71 }
\def\llcorner{\delimiter"4\msafam@78\msafam@78 }
\def\lrcorner{\delimiter"5\msafam@79\msafam@79 }
\def\yen{{\mathhexbox@\msafam@55 }}
\def\checkmark{{\mathhexbox@\msafam@58 }}
\def\circledR{{\mathhexbox@\msafam@72 }}
\def\maltese{{\mathhexbox@\msafam@7A }}

\font\tenmsb=msbm10
\font\sevenmsb=msbm7
\font\fivemsb=msbm5
\newfam\msbfam
\textfont\msbfam=\tenmsb
\scriptfont\msbfam=\sevenmsb
\scriptscriptfont\msbfam=\fivemsb
\edef\msbfam@{\hexnumber@\msbfam}
\def\Bbb#1{\fam\msbfam\relax#1}
\def\widehat#1{\setboxz@h{$\m@th#1$}%
 \ifdim\wdz@>\tw@ em\mathaccent"0\msbfam@5B{#1}%
 \else\mathaccent"0362{#1}\fi}
\def\widetilde#1{\setboxz@h{$\m@th#1$}%
 \ifdim\wdz@>\tw@ em\mathaccent"0\msbfam@5D{#1}%
 \else\mathaccent"0365{#1}\fi}
\font\teneufm=eufm10
\font\seveneufm=eufm7
\font\fiveeufm=eufm5
\newfam\eufmfam
\textfont\eufmfam=\teneufm
\scriptfont\eufmfam=\seveneufm
\scriptscriptfont\eufmfam=\fiveeufm

\let\goth\frak

\catcode`\@=\csname pre amssym.def at\endcsname


\overfullrule0pt

\def\ifundefined#1{\expandafter\ifx\csname#1\endcsname\relax}

\newtoks\sectionnumber
\newcount\equationnumber
\newcount\thnumber
\newcount\refnum
\newcount\exnumber
\def\nextex{\bLP \the\sectionnumber.\the\exnumber \quad
 \advance\exnumber by 1}

\nopagenumbers
\headline={\ifnum\pageno=1 \hfill \else\hss{\tenrm--\folio--}\hss \fi}

\def\assignnumber#1#2{%
	\ifundefined{#1}\relax\else\message{#1 already defined}\fi
	\expandafter\xdef\csname#1\endcsname
 {\if-\the\sectionnumber\else\the\sectionnumber.\fi\the#2}%
	}%

\hyphenation{pre-print}

\let\mPP=\medbreak
\let\bPP=\bigbreak
\def\br={\hfil\break}
\def\LP{\par\noindent}
\def\sLP{\smallbreak\noindent}
\def\mLP{\medbreak\noindent}
\def\bLP{\bigbreak\noindent}

\def\Sec#1 #2 {\vskip0pt plus.1\vsize\penalty-250\vskip0pt plus-.1\vsize
	\bigbreak\bigskip
	\sectionnumber{#1} \equationnumber0\thnumber0
	\noindent{\bf\if-#1\else #1. \fi#2}\par
	\nobreak\smallskip\noindent}
\def\Subsec#1 #2 {\bLP{\bf
 	\if-\the\sectionnumber\else\the\sectionnumber.\fi#1. #2.}\quad}
\def\Proof{\goodbreak\sLP{\bf Proof}\quad}
\def\Ref{\vskip0pt plus.1\vsize\penalty-250\vskip0pt plus-.1\vsize
	\bigbreak\bigskip\leftline{\bf References}\nobreak\smallskip
	\frenchspacing}

\def\wrt{with respect to }
\def\st{such that}

\def\halmos{\hbox{\vrule height0.31cm width0.01cm\vbox{\hrule height
 0.01cm width0.3cm \vskip0.29cm \hrule height 0.01cm width0.3cm}\vrule
 height0.31cm width 0.01cm}}
\def\hhalmos{{\unskip\nobreak\hfil\penalty50
	\quad\vadjust{}\nobreak\hfil\halmos
	\parfillskip=0pt\finalhyphendemerits=0\par}}
\def\nologo{\expandafter\let\csname logo\string @\endcsname=\empty}
\def\:{:\allowbreak }

\def\al{\alpha}
\def\be{\beta}

\def\de{\delta}
\def\ep{\varepsilon}

\def\ka{\kappa}
\def\la{\lambda}
\def\si{\sigma}

\def\De{\Delta}

\def\CC{{\Bbb C}}

\def\RR{{\Bbb R}}

\def\FSA{{\cal A}}
\def\FSB{{\cal B}}

\def\FSH{{\cal H}}

\def\FSL{{\cal L}}

\def\FST{{\cal T}}
\def\FSU{{\cal U}}

\def\gog{{\goth g}}

\def\id{{\rm id}}

\def\tr{{\rm tr}\,}
\def\const{{\rm const.}\,}

\def\Re{{\rm Re}\,}

\def\Span{{\rm Span}}

\def\LHS{left hand side}
\def\RHS{right hand side}
\def\half{{\scriptstyle{1\over2}}}

\def\rep{representation}
\let\pa=\partial
\let\lan=\langle
\let\ran=\rangle
\let\ten=\otimes
\let\wt=\widetilde
\let\wh=\widehat
\let\wb=\overline

\def\eq#1{\relax
	\global\advance\equationnumber by 1
	\assignnumber{EN#1}\equationnumber
	{\rm (\csname EN#1\endcsname)}}
\def\eqtag#1{\relax\ifundefined{EN#1}\message{EN#1 undefined}{\sl (#1)}%
	\else(\csname EN#1\endcsname)\fi%
	}

\def\thname#1{\relax
	\global\advance\thnumber by 1
	\assignnumber{TH#1}\thnumber
	\csname TH#1\endcsname}
\def\Assumption#1 {\bLP{\bf Assumption \thname{#1}}\quad}
\def\Cor#1 {\bLP{\bf Corollary \thname{#1}}\quad}
\def\Def#1 {\bLP{\bf Definition \thname{#1}}\quad}
\def\Example#1 {\bLP{\bf Example \thname{#1}}\quad}
\def\Lemma#1 {\bLP{\bf Lemma \thname{#1}}\quad}
\def\Prop#1 {\bLP{\bf Proposition \thname{#1}}\quad}
\def\Remark#1 {\bLP{\bf Remark \thname{#1}}\quad}
\def\Theor#1 {\bLP{\bf Theorem \thname{#1}}\quad}
\def\thtag#1{\relax\ifundefined{TH#1}\message{TH#1 undefined}{\sl #1}%
	\else\csname TH#1\endcsname\fi}

\def\refitem#1 #2\par{\ifundefined{REF#1}
 \global\advance\refnum by1%
 \expandafter\xdef\csname REF#1\endcsname{\the\refnum}%
 \else\item{\ref{#1}}#2\sLP\fi}
\def\ref#1{\ifundefined{REF#1}\message{REF#1 is undefined}\else
	[\csname REF#1\endcsname]\fi}
\def\reff#1#2{\ifundefined{REF#1}\message{REF#1 is undefined}\else
	[\csname REF#1\endcsname, #2]\fi}
\refnum 0

\font\titlefont=cmr10 scaled \magstep3
\font\rmten=cmr10 at 10truept
\font\ttten=cmtt10 at 10truept
\font\slten=cmsl10 at 10truept
\font\bften=cmbx10 at 10truept

\def\Fun{{\rm Fun}}
\def\Pol{{\rm Pol}}
\def\End{{\rm End}}
\def\Ahat{\wh\FSA}

\centerline{\titlefont
General Compact Quantum Groups, a Tutorial}
\bigskip
\centerline{TOM H. KOORNWINDER}
\bigskip
\midinsert
\narrower
{\rmten \noindent{\bften Abstract}\quad
A tutorial introduction is given to general Hopf algebras and
to general compact quantum groups. In the definition and further treatment
of compact quantum groups C*-algebras are avoided. Contact with
Woronowicz's compact matrix quantum groups is made at a later stage.}
\endinsert

\refitem{Abe}

\refitem{Arv76}

\refitem{Dij94}

\refitem{EfRu93}

\refitem{Haz78}

\refitem{Swe}

\refitem{Tit39}

\refitem{Wan93}

\refitem{Wor87}

\refitem{Wor88}

\refitem{Wor91}

\refitem{Wor93}

\par

\Sec0 {Introduction} \footnote{}
{\rmten
University of Amsterdam, Faculty of Mathematics and Computer Science,
Plantage Muidergracht 24, 1018 TV Amsterdam. The Netherlands,
{\slten email:} {\ttten thk@fwi.uva.nl}}

It is the purpose of this paper to give a tutorial introduction to
Hopf algebras and general compact quantum groups.
An earlier version of the paper was part of the lecture notes
of my course on ``Quantum groups and $q$-special functions''
at the European School of Group Theory 1993, Trento, Italy.
Eventually, this paper may appear as part of a book containing the
notes of the main courses at this School.

For the reader who is new to this subject, it is crucial that he complements
reading of the present paper with the study of papers dealing with
$SU_q(2)$ and other special quantum groups.

I now describe the various sections in some more detail.
Section 1 presents the basic theory of Hopf algebras and of co\rep s.
Here (and in \S2) proofs are either given in full detail or they are sketched
such that the reader can easily fill in the gaps.
No attempt has been made to high mathematical sophistication or big
generality.

Section 2 deals with compact quantum groups.
The original plan was to give here an account of Woronowicz's celebrated
theory \ref{Wor87} of compact matrix quantum groups, but this section grew out
into an alternative approach to compact quantum groups, avoiding
$C^*$-algebras in the definition and in the proofs, but formulating
everything on the Hopf $*$-algebra level.
The $C^*$-algebra completion now appears as a final observation instead
of an essential part of the definition.
Both in \ref{Wor87} and in \S2
the Haar functional plays a crucial role.
The approach of \S2, developed joint with M. S. Dijkhuizen \ref{Dij94},
may be somewhat
shorter and easier to grasp than the $C^*$-algebra approach.
We believe that this approach
is very well suited for application to the special compact matrix
quantum groups most commonly studied nowadays.
The section concludes with a comparison of various approaches to
compact quantum groups which have appeared in the literature.
Particular mention here deserves the paper \ref{EfRu93}
by Effros \& Ruan, who earlier introduced
the same algebras as our CQG algebras, but called them differently and
also developed the theory in a different direction.

I thank Mathijs Dijkhuizen for commenting on
a preliminary version of this paper.

\Sec1 {Generalities about Hopf algebras}
Standard references about Hopf algebras are the books by Abe \ref{Abe}
and Sweedler \ref{Swe},
see also Hazewinkel \reff{Haz78}{\S37.1}.
Below we will assume ground field $\CC$.
By the {\sl tensor product} $V\ten W$ of two linear spaces
$V$ and $W$ we will always mean the {\sl algebraic} tensor product.
Thus the elements of $V\ten W$ are finite sums of elements $v_i\ten w_i$
($v_i\in V$, $w_i\in W$).

\Subsec1 {Hopf algebras}
The reader will be familiar with the concept of
an {\sl associative algebra with unit}
(or shortly an {\sl algebra}), i.e.,
a linear space $\FSA$ with a bilinear
mapping
$(a,b)\mapsto ab\colon\FSA\times\FSA\to\FSA$
and with a special nonzero element $1\in\FSA$ \st
$$
(ab)c=a(bc)\quad{\rm and}\quad 1a=a=a1.
\eqno\eq{1}
$$
When we define the {\sl multiplication} as the linear mapping
$m\colon\FSA\ten\FSA\to\FSA$ \st\ $m(a\ten b)=ab$
and the {\sl unit} as the linear mapping
$\eta\colon\CC\to\FSA$ \st\ $\eta(1)=1_\FSA$,
then we can rephrase \eqtag{1} as
$$
\eqalignno{
&m\circ(m\ten\id)=m\circ(\id\ten m),&\eq{112}
\cr
&m\circ(\eta\ten\id)=\id=m\circ(\id\ten\eta).&\eq{113}
\cr}
$$
Here the two sides of \eqtag{112} are linear mappings from
$\FSA\ten\FSA\ten\FSA$ to $\FSA$, while the three parts of \eqtag{113}
are linear mappings from $\FSA$ to $\FSA$.
In \eqtag{113} we identify $\CC\ten\FSA$ with $\FSA$ by identifying
$c\ten a$ with $ca$ ($c\in\CC$, $a\in\FSA$).
Thus $(\eta\ten\id)(a)=(\eta\ten\id)(1\ten a)=1_\FSA\ten a$.

\Def{2}
A {\sl coassociative coalgebra with counit}
(or shortly a {\sl coalgebra})
is a linear space $\FSA$
with linear mappings
$\De\colon\FSA\to\FSA\ten\FSA$ ({\sl comultiplication})
and
$\ep\colon\FSA\to\CC$ ({\sl counit}) (nonzero) \st
$$
\eqalignno{
&(\De\ten\id)\circ\De=(\id\ten\De)\circ\De\quad
\hbox{({\sl coassociativity}),}
&\eq{2}
\cr
&(\ep\ten\id)\circ\De=\id=(\id\ten\ep)\circ\De.
&\eq{3}
\cr}
$$
Here the two sides of \eqtag{2} are linear mappings from $\FSA$ to
$\FSA\ten\FSA\ten\FSA$, while the three parts of \eqtag{3} are linear
mappings from $\FSA$ to $\FSA$. Here we used again the identification of
$\CC\ten\FSA$ or $\FSA\ten\CC$ with $\FSA$.

\bPP
If $\FSA$ is an algebra with identity element $1$ then $\FSA\ten\FSA$
naturally becomes an algebra with identity element $1\ten1$ if we define
the product of $a\ten b$ and $c\ten d$ as $ac\ten bd$.

When we speak of an algebra homomorphism then we will always mean
that the homomorphism mapping also sends $1$ to $1$
({\sl unital} algebra homomorphism).

\Def{3}
A {\sl bialgebra} is a linear space $\FSA$ which has the structure
of both an algebra and a coalgebra \st\
the mappings $\De\colon\FSA\to\FSA\ten\FSA$ and
$\ep\colon\FSA\to\CC$ are algebra homomorphisms.

\Def{4}
A {\sl Hopf algebra} is a bialgebra $\FSA$ together with a linear
mapping $S\colon\FSA\to\FSA$ ({\sl antipode}) \st
$$
\eqalignno{
&m\circ(S\ten\id)\circ\De=\eta\circ\ep=m\circ(\id\ten S)\circ\De
&\eq{4}
\cr
\noalign{\hbox{(identities of linear mappings from $\FSA$ to $\FSA$), i.e.,}}
&(m\circ(S\ten\id)\circ\De)(a)=\ep(a)\,1=(m\circ(\id\ten S)\circ\De)(a),
\quad a\in\FSA.
&\eq{110}
\cr}
$$

\bPP
If $V$ is a linear space then the {\sl flip operator}
$\si\colon V\ten V\to V\ten V$ will be the linear operator \st\
$\si(v_1\ten v_2)=v_2\ten v_1$ for $v_1,v_2\in V$.
Note that an algebra $\FSA$ is commutative iff
$m\circ\si=m$. By analogy, we define a coalgebra $\FSA$ to be
{\sl cocommutative} if $\si\circ\De=\De$.

\Example{5}
Let $G$ be a group and let $\FSA:=\Fun(G)$ be the algebra (under
pointwise multipication) of all complex-valued functions on $G$.
To some extent, the algebra $\Fun(G\times G)$ of all  complex-valued
functions on $G\times G$ can be viewed as the tensor product
$\FSA\ten\FSA$. Just write
$(a\ten b)(x,y):=a(x)\,b(y)$ if $a,b\in\FSA$ and $x,y\in G$.
However, $\Fun(G\times G)$ is not the algebraic tensor product of
$\Fun(G)$ with $\Fun(G)$ except if $G$ is a finite group.
Let us for the moment not worry about this.
Since $(m(a\ten b))(x)=(ab)(x)=a(x)\,b(x)=(a\ten b)(x,x)$ if
$a,b\in\Fun(G)$, we can write
$$
(m(F))(x)=F(x,x),\quad F\in\Fun(G\times G),\;x\in G.
\eqno\eq{5}
$$
Now define the comultiplication, counit and antipode by
$$
\eqalignno{
&(\De(a))(x,y):=a(xy),\quad a\in\FSA,\;x,y\in G,
&\eq{6}
\cr
&\ep(a):=a(e),\quad a\in\FSA,
&\eq{7}
\cr
&
(S(a))(x):=a(x^{-1}),\quad a\in\FSA,\;x\in G.
&\eq{8}
\cr}
$$
The general philosophy here is that all properties of and
information about the group can be stored in the algebra $\Fun(G)$.
Thus group multiplication, group identity and group inverse are
described on the level of $\Fun(G)$ by $\De$, $\ep$ and $S$,
respectively.

Evidently, $\De$ and $\ep$ are algebra homomorphisms and
$S$ is a linear mapping (in this example also an algebra
homomorphism, while $S^2=\id$).
Furthermore, the group axioms yield the Hopf algebra axioms
\eqtag{2}, \eqtag{3} and \eqtag{4}.
This can be seen for the first identity in \eqtag{110} by observing from
\eqtag{6} and \eqtag{8} that
$(((S\ten\id)\circ\De)(a))(x,y)=a(x^{-1}y)$ and
next from \eqtag{5} that the left hand side of \eqtag{110} evaluated
in $x\in G$ equals
$a(x^{-1}x)$=$a(e)$. By \eqtag{7} the evaluation of the middle part of
\eqtag{110} in $x$ yields the same.

Cocommutativity $\si\circ\De=\De$ would be equivalent here to
$a(xy)=a(yx)$ for all $a\in\Fun(G)$. Thus $\Fun(G)$ is cocommutative
iff the group $G$ is abelian.

Everything above holds rigorously with algebraic tensor products if
$G$ is a finite group.
Now suppose that $G$ is a subgroup of $SL(n,\CC)$
(the group of complex $n\times n$ matrices of determinant 1).
Usually (but not necessarily),
this subgroup $G$ may be thought to be {\sl algebraic},
i.e.\ closed in the Zariski topology.
Let $\Pol(G)$ consist of all complex-valued functions depending on
$x\in G$ which
can be written as polynomials in the matrix elements $x_{ij}$ of $x$.
Let $\Pol(G\times G)$ consist
of all complex-valued functions of $(x,y)\in G\times G$ which can be
written as polynomials in the matrix elements $x_{ij}$ and $y_{ij}$ of $x$
and $y$.
Then $\Pol(G\times G)$ can be identified with $\Pol(G)\ten\Pol(G)$.
Now $\FSA:=\Pol(G)$ becomes a Hopf algebra in the algebraic sense
with the above defined operations.

Let $t_{ij}$ be the element of $\FSA=\Pol(G)$ \st\
$t_{ij}(x)=x_{ij}$ ($x\in G$). Then the $t_{ij}$ form a set of generators
of the algebra $\FSA$. In fact, $\FSA$ is the quotient algebra
obtained when the free abelian algebra with 1 generated by the $t_{ij}$
is divided by the ideal of all elements in this free algebra
which vanish on $G$.
The Hopf algebra operations can now be specified by defining them for the
generators. Thus, since
$$
(\De(t_{ij}))(x,y)=t_{ij}(xy)=(xy)_{ij}=
\sum_{k=1}^n x_{ik}\,y_{kj}=
\sum_{k=1}^n t_{ik}(x)\,t_{kj}(y),
$$
we have
$$
\De(t_{ij})=\sum_{k=1}^n t_{ik}\ten t_{kj},
\eqno\eq{53}
$$
and similarly
$$
\ep(t_{ij})=\de_{ij},
\quad
S(t_{ij})=T_{ji},
\eqno\eq{54}
$$
where $T_{ji}$ is the cofactor of the $(ji)$th entry in the matrix
$(t_{kl})_{k,l=1,\ldots,n}$.

\bPP
The following notation is often useful.
Let $\FSA$ be a coalgebra. If $a\in\FSA$ then we can choose sets of
elements $a_{(1)i}$ and $a_{(2)i}$ in $\FSA$ ($i$ running over a
finite set) \st\
$\De(a)=\sum_i a_{(1)i}\ten a_{(2)i}$.
We write this symbolically as
$$
\De(a)=\sum_{(a)}a_{(1)}\ten a_{(2)},\quad a\in\FSA.\eqno\eq{9b}
$$
Similarly, we write
$$
(\De\ten\id)(\De(a))=
\sum_{(a)}a_{(1)}\ten a_{(2)}\ten a_{(3)},\quad a\in\FSA.
$$
This notation is justified by the coassociativity \eqtag{2}.

Now we can express identities for Hopf algebras which involve
comultiplication by this notation.
For instance, \eqtag{3} (applied to $a\in \FSA$) and \eqtag{110}
can be written as
$$\eqalignno{
&\sum_{(a)}\ep(a_{(1)})\,a_{(2)}=a
=\sum_{(a)}\ep(a_{(2)})\,a_{(1)},
&\eq{111}
\cr
&\sum_{(a)}S(a_{(1)})\,a_{(2)}=\ep(a)\,1=\sum_{(a)}a_{(1)}\,S(a_{(2)}).
&\eq{105}
\cr}
$$

\Prop{7}
Let $\FSA$ be a bialgebra.
If an antipode $S$ exists \st\ $\FSA$ becomes a Hopf algebra then $S$ is
unique.
\Proof
Define a {\sl convolution product} $F*G$ of linear mappings $F$ and $G$ from
the bialgebra $\FSA$ to itself by
$$
(F*G)(a):=(m\circ(F\ten G)\circ\De)(a)=\sum_{(a)} F(a_{(1)})\,G(a_{(2)}),
\quad a\in\FSA.
\eqno\eq{46}
$$
Coassociativity of $\De$ and associativity of $m$ then show that,
under this convolution operation,
$\End(\FSA)$ is an associative algebra.
The properties of counit and unit yield $\eta\circ\ep$ as an identity
element for this algebra.
The antipode property \eqtag{4} can now be interpreted as
$$
S*\id=\eta\circ\ep=\id*S.
\eqno\eq{9a}
$$
Thus, if $S$ exists then it is the two-sided inverse of $\id$ in this
convolution algebra, and therefore unique.\hhalmos

\bPP
Because of \eqtag{9a} we can write the antipode as $S=S*\id*S$.
This can be used
in order to derive further properties of $S$.
In particular, the next proposition states that $S$ is unital, counital,
anti-multiplicative and anti-comultiplicative.

\Prop{8}
Let $\FSA$ be a Hopf algebra. Then, for $a,b\in\FSA$,
$$
\eqalignno{
&S(1)=1,&\eq{10}
\cr
&\ep(S(a))=\ep(a),&\eq{11}
\cr
&S(ab)=S(b)\,S(a),&\eq{12}
\cr
&\De(S(a))=((S\ten S)\circ\si\circ\De)(a)=\sum_{(a)}S(a_{(2)})\ten
S(a_{(1)}).&\eq{13}
\cr}
$$

\Proof
Formula \eqtag{10} follows by putting $a:=1$ in \eqtag{110}, while
\eqtag{11} follows from \eqtag{111} and \eqtag{105}.
For the proof of \eqtag{12} write
$$
\eqalignno{
S(b)\,S(a)=&
\sum_{(a),(b)}S(b_{(1)})\,S(a_{(1)})\,(\ep(a_{(2)}b_{(2)})\,1)
\cr
=&\sum_{(a),(b)} S(b_{(1)})\,S(a_{(1)})\,a_{(2)}\,b_{(2)}\,
S(a_{(3)}b_{(3)})
\cr
=&\sum_{(a),(b)}\ep(a_{(1)})\,\ep(b_{(1)})\,S(a_{(2)}b_{(2)})=S(ab).
\cr}
$$
In the first identity \eqtag{111} was applied twice.
In the second identity we used \eqtag{105} with $a:=a_{(2)}b_{(2)}$
and also the fact that $\De$ is an algebra homomorphism.
In the third identity we used \eqtag{105} twice.
The fourth identity contains two further applications of \eqtag{111}.

The proof of \eqtag{13} is given by
$$
\eqalignno{
\sum_{(a)} S(a_{(2)})\ten S(a_{(1)})=&
\sum_{(a)}\ep(a_{(3)})\,S(a_{(2)})\ten S(a_{(1)})
\cr
=&\sum_{(a)}(S(a_{(2)})\ten S(a_{(1)}))\,
(\ep(a_{(3)})\,1\ten 1)
\cr
=&\sum_{(a)}(S(a_{(2)})\ten S(a_{(1)}))\,
(a_{(3)}\ten a_{(4)})\,\De(S(a_{(5)}))
\cr
=&\sum_{(a)}(S(a_{(2)})\,a_{(3)}\ten S(a_{(1)})\,a_{(4)})\,
\De(S(a_{(5)}))
\cr
=&\sum_{(a)}\ep(a_{(2)})\,(1\ten S(a_{(1)})\,a_{(3)})\,\De(S(a_{(4)}))
\cr
=&\sum_{(a)}(1\ten S(a_{(1)})\,a_{(2)})\,\De(S(a_{(3)})
\cr
=&\sum_{(a)}\ep(a_{(1)})\,\De(S(a_{(2)}))=\De(S(a)).
\cr}
$$
For the third identity above, apply $\De$ to the second identity in
\eqtag{105}.

\Example{6}
Let $\gog$ be a complex
Lie algebra and $\FSA:=\FSU(\gog)$ its universal enveloping
algebra. $\FSU(\gog)$ is defined as the quotient algebra
$\FST(\gog)/J$, where $\FST(\gog)$ is the tensor algebra of $\gog$
and $J$ is the ideal in $\FST(\gog)$ which is generated by the elements
$XY-YX-[X,Y]$ ($X,Y\in\gog$). We define $\De$, $\ep$ and $S$ first on
$\gog$:
$$
\De(X):=X\ten 1+ 1\ten X,\quad
\ep(X):=0,\quad
S(X):=-X\qquad
\hbox{for $X\in\gog$}.
\eqno\eq{9}
$$
On $\CC$ we declare the operators to be unital.
Now we can check the coalgebra axioms and the antipode axiom already
for the operators acting on $\CC\oplus\gog$.
In fact, the counit and antipode axiom force $\ep$ and $S$ to be as in
\eqtag{9}.
Next we extend $\De$ and $\ep$ to $\FST(\gog)$ as algebra homomorphisms
and $S$ as anti-algebra homomorphism.
Now check that $\De(J)\i\FST(\gog)\ten J+J\ten\FST(\gog)$,
$\ep(J)=0$ and $S(J)\i J$.
This allows us to consider $\De$, $\ep$ and $S$ as
operators on $\FSU(\gog)$. Finally we have to check the Hopf algebra
axioms on all of $\FSU(\gog)$, by using that they are already satisfied
on a subspace of generators.
Note that $\FSA$ is cocommutative, but generally not commutative,
unless $\gog$ is an abelian Lie algebra.
Note that the antipode satisfies $S^2=\id$.

\Remark{6a}
A further motivation for the concept of comultiplication is the wish
to construct tensor products of \rep s of algebras.
Suppose $\FSA$ is an algebra.
Let $\pi_1$ and $\pi_2$ be algebra \rep s of $\FSA$ on finite
dimensional linear spaces
$V_1$ and $V_2$ resp., i.e., algebra homomorphisms of $\FSA$
to the algebras of linear endomorphisms of $V_1$ and $V_2$ respectively.
Then $a_1\ten a_2\mapsto \pi_1(a_1)\ten\pi_2(a_2)\colon
\FSA\ten\FSA\to\End(V_1\ten V_2)$ is a \rep\ of $\FSA\ten\FSA$
on $V_1\ten V_2$. In order to obtain from this \rep\ a \rep\ of
$\FSA$ on $V_1\ten V_2$, we need an algebra homomorphism from
$\FSA$ to $\FSA\ten\FSA$.  Without further structure on $\FSA$ there is no
canonical method  for this. However, if $\FSA$ is a Hopf algebra, the
desired mapping is provided by the comultiplication
$\De$. Then we can define the tensor product \rep\ $\pi_1\ten\pi_2$
of $\FSA$ as the composition
$(\pi_1\ten\pi_2)\circ\De$.
Associativity of this tensor product is precisely assured by
the coassociativity \eqtag{2} of $\De$.

Other Hopf algebra axioms are also meaningful in this context.
For instance, the counit $\ep$ gives a one-dimensional algebra \rep\ of
$\FSA$. Then the counit axiom \eqtag{3} implies that, for each finite
dimensional algebra \rep\ $\pi$ of $\FSA$, we have that
$\pi\ten\ep=\pi=\ep\ten\pi$.
Also, if $\pi$ is an algebra \rep\ of $\FSA$ on a finite dimensional
linear space $V$ and if $V^*$ is the linear dual of $V$, then we can
define an algebra \rep\ $\pi^*$ of $\FSA$ on $V^*$ by
$$
\lan \pi^*(a)\,v^*,v\ran:=\lan v^*,\pi(S(a))\,v\ran,\quad
v\in V,\quad v^*\in V^*.
$$

Recall that the tensor  product of two \rep s $\pi_1$ and $\pi_2$ of
a Lie algebra $\gog$ is provided by
$((\pi_1\ten\pi_2)(X))(v_1\ten v_2)=(\pi_1(X)\,v_1)\ten v_2+
v_1\ten(\pi_2(X)\,v_2)$. Compare this with the comultiplication for
$\gog$ as given by \eqtag{9}.

\bPP
One should be aware that the antipode in a Hopf algebra $\FSA$
is not necessarily invertible as a linear mapping from $\FSA$ to itself.

We may consider $\FSA$ also as a Hopf algebra with opposite
multiplication $m\circ\si$ and opposite comultiplication
$\si\circ\De$, and $\ep$, $\eta$ and $S$ unchanged.
Then $S$ is a Hopf algebra homomorphism from $\FSA$ with the old
Hopf algebra structure to $\FSA$ with the new one.
However, if we change only one of the two operations of multiplication
and comultiplication into its opposite then we still have a bialgebra,
but this will be a Hopf algebra if and only if $S$ is invertible.
The new antipode
will then be precisely $S^{-1}$. Indeed, let $\FSA'$ be $\FSA$ equipped
with opposite multiplication and suppose $\FSA'$ is a Hopf algebra
with antipode $S'$. Now apply $S$ to
$$
\sum_{(a)}a_{(2)}\,S'(a_{(1)})=\ep(a)1=
\sum_{(a)}S'(a_{(2)})\,a_{(1)}.
$$
This yields that $SS'$ is the two-sided inverse of $S$ under convolution.
Hence $SS'=\id$, and also $S'S=\id$ by reverting the roles of $S$ and
$S'$. Conversely, if $S$ is invertible, then define $S':=S^{-1}$ and show
that $S'$ is an antipode for $\FSA'$.
\bPP
Recall that an (associative) algebra $\FSA$ (with 1) is a {\sl $*$-algebra}
if there is a mapping $a\mapsto a^*\colon\FSA\to\FSA$ (an {\sl
involution}) \st\
$(a^*)^*=a$, $(a+b)^*=a^*+b^*$, $(\la a)^*=\wb\la\,a^*$,
$(ab)^*=b^*a^*$, $1^*=1$.  If $\FSA$ is a $*$-algebra then
$\FSA\ten\FSA$ becomes a $*$-algebra with $(a\ten b)^*:=a^*\ten b^*$.
If $\FSA$ and $\FSB$ are $*$-algebras then $F\colon\FSA\to\FSB$ is called
a {\sl $*$-homomorphism} if $F$ is an algebra homomorphism \st\
$F(a^*)=(F(a))^*$.

\Def{9}
A Hopf $*$-algebra is a Hopf algebra $\FSA$ which, as an algebra,
is also a $*$-algebra \st\ $\De\colon\FSA\to\FSA\ten\FSA$ and
$\ep\colon\FSA\to\CC$ are $*$-homomorphisms.

\bPP
Originally, the property stated in the next Proposition was part of the
definition of a Hopf $*$-algebra. I thank S. Zakrzewski for helpful
correspondence about this.

\Prop{10}
In a Hopf $*$-algebra $\FSA$ we have
$$
S\circ*\circ S\circ *=\id.
\eqno\eq{43}
$$
In particular, $S$ is invertible with inverse $*\circ S\circ *$.
\Proof
Apply \eqtag{4} to $a^*$ and next apply the involution to all members of
\eqtag{4}. This shows that $*\circ S\circ *$ is an antipode for $\FSA'$,
i.e., for
$\FSA$ with opposite multiplication. Hence
$*\circ S\circ *=S^{-1}$.\hhalmos

\Example{62}
Let $G$ be a group and let $\FSA:=\Fun(G)$ as in Example \thtag{5}
For $f\in\Fun(G)$ define $f^*(x):=\wb{f(x)}$ ($x\in G$).
Similarly, for $F\in\Fun(G\times G)$ define $F^*(x,y):=\wb{F(x,y)}$
($x,y\in G$). Then $\Fun(G)$ and $\Fun(G\times G)$ are commutative
$*$-algebras and $\De\colon\Fun(G)\to\Fun(G\times G)$ and
$\ep\colon\Fun(G)\to\CC$ are $*$-homomorphisms.
So, apart from the fact that $\De$ does not map $\Fun(G)$ into
$\Fun(G)\ten\Fun(G)$, we have $\Fun(G)$ as an example of a
Hopf $*$-algebra.

\Example{11}
Let $G$ be a complex Lie group given as a closed connected subgroup of
$SL(n,\CC)$ and let $G_0$ be a real connected Lie group and a real
form of $G$. Then every $a\in\Pol(G)$ is completely determined by
its restriction to $G_0$. Suppose that, for each $a\in\Pol(G)$,
there exists $a^*\in\Pol(G)$ \st\ $a^*(x)=\wb{a(x)}$ for $x\in G_0$.
Then the Hopf algebra $\Pol(G)$ becomes a Hopf $*$-algebra with this
mapping $*$. Conversely, if the Hopf algebra $\Pol(G)$ is a Hopf $*$-algebra
then we may define a real form $G_0$ of $G$ by $G_0:=\{x\in G\mid
\wb{a(x)}=a^*(x)\quad\forall a\in\Pol(G)\}$.

Thus, on the level of polynomial
function algebras, a real Lie group is described
by the algebra of functions on its complexification together with an
involution for this algebra.

In particular, let the $t_{ij}$ (cf.\ Example \thtag{5})
be the generators of the algebra
$\FSA=\Pol(G)$. If we suppose that
$G_0\i SL(n,\RR)$ then $t_{ij}^*(x)=\wb{t_{ij}(x)}=t_{ij}(x)$ for
$x\in G_0$. Hence $t_{ij}^*=t_{ij}$.
This gives the action of $*$ on the generators and next, by anti-linear
homomorphic continuation, on all of $\FSA$.
However, if $G_0\i SU(n)$ then
$t_{ij}^*(x)=\wb{t_{ij}(x)}=t_{ji}(x^{-1})=(S(t_{ji}))(x)$ if $x\in G_0$.
Hence $t_{ij}^*=S(t_{ji})$. Again, the involution on all of $\FSA$ follows
by anti-linear homomorphic continuation.

\Subsec2 {Duality for Hopf algebras}
Let $\FSA$ be a Hopf algebra
and let $\FSA^*$ be its algebraic linear dual, i.e., the space of all linear
mappings
$f\colon \FSA\to\CC$. We will write $\lan f,a\ran:=f(a)$
($f\in\FSA^*$, $a\in\FSA$).
The algebraic tensor product $\FSA^*\ten\FSA^*$ is a subspace of
$(\FSA\ten\FSA)^*$ by the rule
$$
\lan f\ten g,a\ten b\ran=
\lan f,a\ran\,\lan g,b\ran,\quad
f,g\in\FSA^*,\;a,b\in\FSA.
$$
Unless $\FSA$ is finite dimensional, this will be a proper subspace.
By duality, the Hopf algebra operations on
$\FSA$ can be transferred to $\FSA^*$. We define
$$
\eqalignno{
&\lan fg,a\ran:=\lan f\ten g,\De(a)\ran
=\textstyle\sum_{(a)}\lan f,a_{(1)}\ran\,\lan g,a_{(2)}\ran,&\eq{38}
\cr
&\lan\De(f),a\ten b\ran:=\lan f,ab\ran,&\eq{39}
\cr
&\lan 1_{\FSA^*},a\ran:=\ep_\FSA(a),&\eq{108}
\cr
&\ep_{\FSA^*}(f):=\lan f,1_\FSA\ran,&\eq{109}
\cr
&\lan S(f),a\ran:=\lan f,S(a)\ran.&\eq{61}
\cr}
$$
However, $\De(f)$, as defined by \eqtag{39}, will be an element of
$(\FSA\ten\FSA)^*$ and not necessarily of
$\FSA^*\ten\FSA^*$. Still, with a suitable adaptation of the
definition of Hopf algebra, the Hopf algebra
axioms for $\FSA^*$ with the above operations can be verified in a
straightforward way.

If $\FSA$ is moreover a Hopf $*$-algebra then we can define an involution
on $\FSA^*$ by
$$
f^*(a):=\wb{\lan f,(S(a))^*\ran}.
\eqno\eq{44}
$$
Note that \eqtag{43} ensures that $(f^*)^*=f$. The antipode is needed in
\eqtag{44} in order to ensure that
$(fg)^*=g^*f^*$. (The more simple definition
$f^*(a):=\wb{\lan f,a^*\ran}$ would make the involution on $\FSA^*$
multiplicative
rather than anti-multiplicative.) The reader should verify that $\FSA^*$
thus indeed becomes a
Hopf $*$-algebra, with suitably adapted definition because of
the fact that $\De$ does not necessarily map
into $\FSA^*\ten\FSA^*$.

Let $\FSA$ be a Hopf algebra. We define left and right
algebra actions of $\FSA^*$ on $\FSA$:
$$
\eqalignno{
&f.a:=(\id\ten f)(\De(a))=\sum_{(a)}f(a_{(2)})\,a_{(1)},&\eq{68}
\cr
&a.f:=(f\ten\id)(\De(a))=\sum_{(a)}f(a_{(1)})\,a_{(2)}.&\eq{69}
\cr}
$$
It is indeed an easy exercise to verify that
$$
(fg).a=f.(g.a),\quad a.(fg)=(a.f).g,\quad
1_{\FSA^*}.a=a.1_{\FSA^*}=a.
$$
An important property is the following.
If $f\in\FSA^*$ \st\ $\De(f)\in\FSA^*\ten\FSA^*$
and if we write $\De(f)=\sum_{(f)}f_{(1)}\ten f_{(2)}$ then
$$
\eqalignno{
&f.(ab)=\sum_{(f)} (f_{(1)}.a)\,(f_{(2)}.b),\quad a,b\in\FSA,&\eq{106}
\cr
&(ab).f=\sum_{(f)} (a.f_{(1)})\,(b.f_{(2)}),\quad a,b\in\FSA.&\eq{107}
\cr}
$$
We give the proof of \eqtag{106}; the proof of \eqtag{107} is similar.
$$
\eqalignno{
f.(ab)=\sum_{(a),(b)}a_{(1)}\,b_{(1)}\,f(a_{(2)}b_{(2)})=
\sum_{(a),(b),(f)}a_{(1)}\,f_{(1)}(a_{(2)})\,b_{(1)}\,f_{(2)}(b_{(2)})&
\cr
=\sum_{(f)}(f_{(1)}.a)\;(f_{(2)}.b).&
\cr}
$$

\Def{60}
Two Hopf algebras $\FSU$ and $\FSA$ are said to be {\sl Hopf algebras
in duality} if there is a bilinear
mapping $(u,a)\mapsto\lan u,a\ran\colon\FSU\times\FSA\to\CC$ \st\
\eqtag{38}--\eqtag{61} are satisfied when $\FSU$ is read
instead of $\FSA^*$.

If $\FSU$ and $\FSA$ are moreover Hopf $*$-algebras and if \eqtag{44} also
holds then we speak about {\sl Hopf $*$-algebras in duality}.

Two Hopf ($*$-)algebras in duality $\FSA$ and $\FSU$ are said to be
{\sl Hopf ($*$-)algebras in nondegenerate duality}
(in fact in doubly nondegenerate duality)
if the two following implications moreover hold:
(i) $(\forall a\in\FSA$ $\lan u,a\ran=0)$ $\Longrightarrow$ $u=0$, and
(ii) $(\forall u\in\FSU$ $\lan u,a\ran=0)$ $\Longrightarrow$ $a=0$.

\bPP
If $\FSU$ and $\FSA$ are Hopf algebras in duality (not necessarily
nondegenerate) then left and right actions of $\FSU$ on $\FSA$ as in
\eqtag{68} and \eqtag{69} can still be defined.

If $\FSA$ is a Hopf ($*$-)algebra and if $\FSU$ is a Hopf
($*$-)subalgebra of $\FSA^*$, i.e., if $\FSU$ is a unital subalgebra
of $\FSA^*$ \st\ $\De(\FSU)\i\FSU\ten\FSU$, $S(\FSU)\i\FSU$ and,
in case of a Hopf $*$-subalgebra, $\FSU^*\i\FSU$, then $\FSU$ and $\FSA$
are obviously Hopf ($*$-)algebras in duality, while the duality is already
nondegenerate on one side: if $u\in\FSU$ and if
$\lan u,a\ran=0$ for all $a\in\FSA$ then $u=0$.

If $\FSU$ and $\FSA$ are Hopf ($*$-)algebras in nondegenerate duality
then $\FSU$ can be viewed as a Hopf ($*$-)subalgebra of $\FSA^*$ and
$\FSA$ as a Hopf ($*$-)subalgebra of $\FSU^*$. So we can write
$\lan u,a\ran=u(a)=a(u)$ for $u\in\FSU$ and $a\in\FSA$.

\Example{61}
Let $G$ be a complex Lie group given as a closed connected subgroup of
$SL(n,\CC)$ and let $\gog$ be its (complex) Lie algebra and
$\FSU:=\FSU(\gog)$
the universal enveloping algebra of $\gog$. Then
$\FSA:=\Pol(G)$ and $\FSU$ are Hopf algebras (cf.\ Examples
\thtag{5} and \thtag{6}).
These Hopf algebras are naturally in duality with each other.
For $X\in\gog$ and $a\in\Pol(G)$ the pairing is given by
$$
\lan X,a\ran={d\over dt}\Bigr|_{t=0}\,a(\exp(tX)).
\eqno\eq{88}
$$
Then, for $f:=X\in\gog$, equations \eqtag{39}--\eqtag{61}
are satisfied in view of \eqtag{88} and \eqtag{9}.

For the left and right actions of $X\in\gog$ on $a\in\Pol(G)$
(cf.\ \eqtag{68} and \eqtag{69}) we then obtain
$$
(X.a)(x)={d\over dt}\Bigr|_{t=0}\,a(x\,\exp(tX)),\quad
(a.X)(x)={d\over dt}\Bigr|_{t=0}\,a(\exp(tX)\,x).
$$
If $X:=X_1X_2\ldots X_k\in\FSU$, with $X_1,\ldots,X_k\in\gog$, then
$$
\eqalignno{
&(X.a)(x)={\pa^k\over\pa t_1\ldots\pa t_k}\Bigr|_{t_1,\ldots,t_k=0}\,
a(x\,\exp(t_1X_1)\ldots\exp(t_kX_k)),\cr
&(a.X)(x)={\pa^k\over\pa t_1\ldots\pa t_k}\Bigr|_{t_1,\ldots,t_k=0}\,
a(\exp(t_1X_1)\ldots\exp(t_kX_k)\,x).
\cr}
$$
So we get the familiar left or right action of $\FSU(\gog)$ on smooth
functions on $G$ by left or right invariant differential operators.
In particular, either of the two last formulas, when evaluated at $x:=e$,
yields the pairing $\lan X_1\ldots X_k,a\ran$.
The formula thus obtained is compatible with \eqtag{88} and \eqtag{38}.
The double non-degeneracy of the pairing can be easily verified.

\Example{51}
Let $G$, $\gog$, $\FSA$ and $\FSU$ be as in the previous example.
Let the connected real Lie group $G_0$ be a real form of $G$
and let $*$ be the
corresponding involution on $\FSA$, by which $\FSA$ becomes a Hopf
$*$-algebra (cf.\ Example \thtag{11}).
Let $\gog_0$ be the (real) Lie algebra of $G_0$. It is a real form
of the complex Lie algebra $\gog$.
Let us compute $X^*$ (as defined by \eqtag{44}) for $X\in\gog_0$
and let us see if $X^*\in\FSU$. For $a\in\FSA$ we have
$$
\lan X^*,a\ran=
\wb{\lan X,(S(a))^*\ran}=
\wb{{d\over dt}\Bigr|_{t=0}\,\wb{a(\exp(-tX))}}=
-\lan X,a\ran.
$$
Here we used that $a^*(x)=\wb{a(x)}$ if $x\in G_0$ (cf. Example
\thtag{11}).
Thus $X^*=-X$ for $X\in\gog_0$. In particular, $*$ maps the Lie
algebra $\gog_0$ to
itself. It follows that $*$ maps $\FSU$ to itself, so $\FSU$ is a
Hopf $*$-algebra in duality with $\FSA$.

Note that in the case $G_0\i SU(n)$ we have $\gog_0\i su(n)$, the
Lie algebra of skew-hermitian matrices. So then the $*$ on $\gog$
induced by the $*$-structure on $\FSA$ has the same effect as taking
the adjoint of the matrix $X\in\gog$. But this is not necessarily true
if $G_0\i SL(n,\RR)$.

\Subsec3 {Co\rep s}

\Def{12}
Let $\FSA$ be a coalgebra. A {\sl corepresentation} of $\FSA$ on
a complex vector space $V$ is defined as a linear mapping
$t\colon V\to V\ten\FSA$ \st\
$$
(t\ten\id)\circ t=(\id\ten\De)\circ t,\quad
(\id\ten\ep)\circ t=\id.
\eqno\eq{13a}
$$

Suppose that $V$ is a finite dimensional vector space with basis
$e_1,\ldots,e_n$.
Then $t$ determines elements $t_{ij}$ of $\FSA$ \st\
$$
t(e_j)=\sum_{i} e_i\ten t_{ij}.
\eqno\eq{14}
$$
Then
$$
\sum_i e_i\ten\De(t_{ik})=
(\id\ten\De)(t(e_k))=
(t\ten\id)(t(e_k))=
\sum_{j,k}e_i\ten t_{ij}\ten t_{jk}
$$
and
$$
\sum_i \ep(t_{ij})\,e_i=
(\id\ten\ep)(t(e_j))=e_j.
$$
Hence
$$
\De(t_{ij})=\sum_k t_{ik}\ten t_{kj},\quad
\ep(t_{ij})=\de_{ij}.
\eqno\eq{15}
$$
Conversely, if elements $t_{ij}$ of $\FSA$ satisfy \eqtag{15}
and if $t$ is defined by \eqtag{14} on the vector space having the
$e_i$ as a basis then $t$ is a co\rep\ of $\FSA$ according to
Definition \thtag{12}.
We call a matrix $(t_{ij})_{i,j=1,\ldots,n}$ satisfying \eqtag{15}
a {\sl matrix corepresentation} of $\FSA$.

\Example{13}
Let $G$ be a group and $\pi$ a \rep\ of $G$ on some vector space $V$.
Let $\Fun(G)$ be as in Example \thtag{5}.
To some extent, we can identify $V\ten\Fun(G)$ with
$\Fun(G;V)$ (the space of $V$-valued functions on $G$).
Just write $(v\ten a)(x):=a(x)\,v$ if $v\in V$, $a\in\FSA$, $x\in G$.
This identification holds in the sense of algebraic tensor products
if $V$ is finite dimensional.
Similarly, identify $V\ten\Fun(G)\ten\Fun(G)$ with
$\Fun(G\times G;V)$ by
$(v\ten a\ten b)(x,y):=a(x)\,b(y)\,v$.
Define $t(v)\in V\ten\Fun(G)$ ($v\in V$)
as the $V$-valued function given by
$$
(t(v))(x):=\pi(x)v,\quad x\in G.
$$
Then $t$ is a co\rep\ of $\Fun(G)$ on $V$.
For this purpose, check equations \eqtag{13a}.
For instance, for the proof of the first identity of \eqtag{13a}
we can show that
$$
(((t\ten\id)\circ t)(v))(x,y)=\pi(x)\,\pi(y)\,v\quad{\rm and}\quad
(((\id\ten\De)\circ t)(v))(x,y)=\pi(xy)\,v.
$$
Let us show the first of these two identities. Fix $v$.
Then there are $v_j\in V$ and $a_j\in\Fun(G)$ \st\
$(t(v))(x)=\pi(x)v=\sum_j a_j(x)\,v_j$, so
$t(v)=\sum_j v_j\ten a_j$.
Then
$$
((t\ten\id)\circ t)(v)=\sum_j (t\ten\id)(v_j\ten a_j)=
\sum_j t(v_j)\ten a_j.
$$
Hence $(((t\ten\id)\circ t)(v))(x,y)$
$$
\qquad=
\sum_j(t(v_j)\ten a_j)(x,y)=
\sum_j a_j(y)\,(t(v_j))(x)=
\sum_j a_j(y)\,\pi(x)\,v_j=\pi(x)\,\pi(y)\,v.
$$

\Example{58}
If $t$ is a matrix \rep\ of a group $G$ then the matrix elements
$t_{ij}$ can be viewed as elements of $\Fun(G)$. They satisfy
$$
t_{ij}(xy)=\sum_k t_{ik}(x)\,t_{kj}(y),\quad
t_{ij}(e)=\de_{ij}.
$$
Hence, by \eqtag{15}, $t$ is a matrix co\rep\ of $\Fun(G)$.
If $t$ is a matrix \rep\ of a group $G\i SL(n,\CC)$  and if the $t_{ij}$
are in $\Pol(G)$ then $t$ is a matrix co\rep\ of $\Pol(G)$.
This holds in particular for the natural matrix \rep\ of $G$ defined by
$t_{ij}(x):=x_{ij}$ ($x\in G$), cf.\ the end of Example \thtag{5}.

\bPP
If $t$ is a matrix co\rep\ of a Hopf algebra $\FSA$ then combination of
\eqtag{15} and \eqtag{4} shows that
$$
\sum_k S(t_{ik})\,t_{kj}=\de_{ij}\,1=
\sum_k t_{ik}\, S(t_{kj}).
\eqno\eq{15a}
$$

\bPP
For a co\rep\ $t$ of $\FSA$ on $V$ we may use a symbolic notation analogous
to \eqtag{9b}:
$$
t(v)=\sum_{(v)} v_{(1)}\ten v_{(2)},\quad v\in V,
\eqno\eq{15b}
$$
which means that we have a sum of elements
$t_{(1)i}\ten t_{(2)i}$, where $t_{(1)i}\in V$ and $t_{(2)i}\in\FSA$.
Then \eqtag{13a} allows us to combine \eqtag{15b} with \eqtag{9b} and to
write
$$
((t\ten\id)\circ t)(v)=\sum_{(v)}v_{(1)}\ten v_{(2)}\ten v_{(3)}=
((\id\ten\De)\circ t)(v).
$$

\Def{14}
Let $t$ be a co\rep\ of a Hopf $*$-algebra $\FSA$ on a vector space
$V$ with hermitian inner product $(\,.\,,\,.\,)$. We call the co\rep\
$t$ {\sl unitary} if
$$
\sum_{(v),(w)}(v_{(1)},w_{(1)})\,w_{(2)}^*\,v_{(2)}=(v,w)\,1,\quad
v,w\in V.
\eqno\eq{16}
$$

\bPP
Condition \eqtag{16} can be equivalently stated as
$$
\sum_{(v)} (v_{(1)},w)\,S(v_{(2)})=
\sum_{(w)} (v,w_{(1)})\,w_{(2)}^*,\quad v,w\in V.
\eqno\eq{17}
$$
We prove the implication \eqtag{16}$\Rightarrow$\eqtag{17},
and leave the other direction to the reader.
Indeed,
$$
\eqalignno{
\sum_{(v)} (v_{(1)},w)\,S(v_{(2)})=&
\sum_{(v),(w)}(v_{(1)},w_{(1)})\,w_{(2)}^*\,v_{(2)}\,S(v_{(3)})
\cr
=&\sum_{(v),(w)}(v_{(1)},w_{(1)})\,w_{(2)}^*\,\ep(v_{(2)})
=\sum_{(w)} (v,w_{(1)})\,w_{(2)}^*.
\cr}
$$

\bPP
If $V$ is finite dimensional with an orthonormal basis $e_1,\ldots,e_n$
then \eqtag{16} implies for the corresponding matrix co\rep\
$(t_{ij})$ that
$$
\sum_i t_{il}^*\,t_{ik}
=\sum_{i,j}(e_i,e_j)\,t_{jl}^*\,t_{ik}
=(e_k,e_l)\,1=\de_{kl}\,1.
$$
In combination with \eqtag{15a} we readily obtain:

\Prop{15}
Let $(t_{ij})$ be a matrix co\rep\ of a Hopf $*$-algebra $\FSA$ and let
$t$ be the corresponding co\rep\ of $\FSA$ on the inner product space
which has $e_1,\ldots,e_n$ as an orthonormal basis.
Then the following conditions are equivalent.
\item{(a)}
$t$ is a unitary co\rep.
\item{(b)}
$\sum_k t_{ki}^*\,t_{kj}=\de_{ij}\,1$.
\item{(c)}
$S(t_{ij})=t_{ji}^*$.
\item{(d)}
$\sum_k t_{ik}\,t_{jk}^*=\de_{ij}\,1$.

\bPP
In Example \thtag{58} we viewed a matrix \rep\ $t$ of a group $G$
as a matrix co\rep\ of $\Fun(G)$. If we have a $*$-operation
on $\Fun(G)$ as in Example \thtag{62}, then it follows from
Proposition \thtag{15} that $t$ is unitary as a matrix \rep\ of $G$
iff it is unitary as a matrix co\rep\ of $\Fun(G)$.

If $\pi$ is a unitary \rep\ of $G$ on an inner product space $V$
and if $t$ is the co\rep\ of $\Fun(G)$ associated with $\pi$
acording to Example \thtag{13} then equations \eqtag{16} and
\eqtag{17} are rewritten versions of
$$
(\pi(x)\,v,\pi(x)\,w)=(v,w)\quad{\rm and}\quad
(\pi(x^{-1})\,v,w)=(v,\pi(x)\,w), \quad x\in G_0.
$$

\Remark{59}
If $t$ is a co\rep\ of $\FSA$ on a vector space $V$ then $t$ gives rise to
an algebra \rep\ $\pi$ of $\FSA^*$ on $V$:
$$
\pi(f)\,v:=(\id\ten f)(t(v))=\sum_{(v)} f(v_{(2)})\,v_{(1)}.
\eqno\eq{86}
$$
Suppose $V$ is finite dimensional with basis $e_1,\ldots,e_n$ and let
the corresponding matrix elements $t_{ij}$ be given by \eqtag{14}.
Combination with \eqtag{86} gives that
$$
\pi(f)\,e_j=\sum_i f(t_{ij})\,e_i.
$$
Hence $\pi$ can then be written as a matrix \rep\
$\pi\colon f\mapsto(\pi_{ij}(f))$ of $\FSA$ with
$\pi_{ij}(f):=f(t_{ij})$.

If $t$ is a unitary matrix co\rep\ of a Hopf $*$-algebra $\FSA$ and if
$\pi$ is the corresponding matrix \rep\ of $\FSA^*$ then $\pi$ is a
$*$-\rep. Indeed,
$$
\pi_{ij}(f^*)=f^*(t_{ij})=\wb{f((S(t_{ij}))^*)}=
\wb{f(t_{ji})}=\wb{\pi_{ji}(f)}.
$$

\Def{16}
Let $t$ be a co\rep\ of a coalgebra $\FSA$ on a vector space $V$.
\item{(a)}
A linear subspace $W$ of $V$ is called {\sl invariant} if
$t(W)\i W\ten\FSA$.
\item{(b)}
The co\rep\ $t$ is called {\sl irreducible} if $V$ and $\{0\}$ are the
only invariant subspaces of $V$
\item{(c)}
Let $s$ be another co\rep\ of $\FSA$ on a vector space $W$.
A linear operator $L\colon V\to W$ is called an
{\sl intertwining operator} for $t$ and $s$ if
$$
s\circ L=(L\ten\id)\circ t.
\eqno\eq{18}
$$
The co\rep s $t$ and $s$ are called {\sl equivalent} if there exists
a bijective intertwining operator for $t$ and $s$.

\bPP
Observe that the relation of equivalence between co\rep s is an
equivalence relation.
One easily checks that two matrix co\rep s $(t_{ij})$ and $(s_{ij})$
of a coalgebra $\FSA$ are equivalent iff there is an invertible square
complex matrix $L$ \st\ $Lt=sL$ (where the products are matrix products).
Furthermore, the matrix co\rep\ $t$ is irreducible iff
$t$ is not equivalent to a matrix co\rep\ of block form
$\pmatrix{*&*\cr0&*\cr}$.

\Prop{17}
Let $t$ be a unitary co\rep\ of a Hopf $*$-algebra on a finite
dimensional Hilbert space $V$.
\item{(a)} Let $W$ be an invariant subspace of
$V$. Then the orthoplement of $W$ in $V$ is also invariant.
\item{(b)} $V$ is a direct sum of invariant subspaces on each of which
the restriction of $t$ is an irreducible co\rep\ of $\FSA$.
\Proof
For the proof of (a) assume that $v$ is orthogonal to $W$.
Write $t(v)=\sum_{(v)}v_{(1)}\ten v_{(2)}$ \st\ the
$v_{(2)}$ are linearly independent. Apply \eqtag{17} with
$w\in W$. Then, by invariance of $W$, the \RHS\ of
\eqtag{17} will be 0. Hence the \LHS\ will be 0, so
$\sum_{(v)}(v_{(1)},w)\,v_{(2)}=0$. Thus the $(v_{(1)},w)$ will be zero.
So the $v_{(1)}$ will be in the orthoplement of $W$.
Part (b) follows by iteration of (a).\hhalmos

\Lemma{18}
Let $L\colon V\to W$ be an intertwining operator for co\rep s $t$ and
$s$ of a coalgebra $\FSA$ on finite dimensional spaces $V$ resp.\ $W$.
\item{(a)}
The image $L(V)$ and the null space $L^{-1}(0)$ are invariant
subspaces of $W$ and $V$, respectively.
\item{(b)} ({\sl first Schur lemma})\quad
If $t$ and $s$ are irreducible then $L=0$ or $L$ is bijective.
\item{(c)} ({\sl second Schur lemma})\quad
If $V=W$, $t=s$ and $t$ is irreducible then $L=\la I$ for some complex
$\la$.

\Proof
In (a)
the invariance of $L(V)$ follows immediately from \eqtag{18}.
In order to prove the invariance of $L^{-1}(0)$,
let $Lv=0$. Then, by \eqtag{18} and \eqtag{15b},
$$
0=(L\ten\id)(t(v))=\sum_{(v)}L(v_{(1)})\ten v_{(2)}.
$$
Now choose the $v_{(2)}$ linearly independent. Then the $L(v_{(1)})$
will be 0. Hence, the $v_{(1)}$ in \eqtag{15b} will belong to $L^{-1}(0)$.
Parts (b) and (c) now follow easily, completely analogous to the proof
of the classical Schur lemmas.\hhalmos

\Lemma{19}
Let $\FSA$ be a coalgebra. Let $t$ be a co\rep\ of $\FSA$ on a finite
dimensional vector space $V$.
Suppose that $V$ is a direct sum of subspaces $V_i$ ($i=1,\ldots,n$)
and that each $V_i$ is a direct sum of subspaces $W_{ij}$
($j=1,\ldots,m_i$) and that there are mutually inequivalent
irreducible co\rep s $t_1,\ldots,t_n$ \st\
each subspace $W_{ij}$ is invariant and $t$ restricted to $W_{ij}$
is equivalent to $t_i$.
Let $U$ be a nonzero invariant subspace of $V$ \st\
$t$ restricted to $U$ is an irreducible co\rep\ $s$.
Then, for some $i$, $U\subset V_i$ and $s$ is equivalent to $t_i$.

\Proof
Let $P_{ij}\colon V\to W_{ij}$ be the projection operator which is
identity on $W_{ij}$ and 0 on the other $W_{kl}$.
Let $\pi_{ij}$ be the restriction of $P_{ij}$ to $U$.
Then $\pi_{ij}\colon U\to W_{ij}$ is an intertwining operator for
$s$ and $t_i$. Hence, by Lemma \thtag{18}(b), it is either 0 or bijective.
Hence, if $\pi_{ij}$ and $\pi_{kl}$ are nonzero, then $t_i$,
$s$ and $t_k$ must be equivalent co\rep s, so $i=k$.
Hence, there is an $i$ \st\
$$
U\i\bigoplus_{k,l}\pi_{kl}(U)=
\bigoplus_{\textstyle{k,l\atop\pi_{kl}\ne0}}W_{kl}\i
\bigoplus_{j=1}^{m_i}W_{ij}=V_i
$$
and $s$ is equivalent to $t_i$.\hhalmos

\bPP
If $\FSA$ and $\FSB$ are coalgebras then $\FSA\ten\FSB$ becomes a
coalgebra with comultiplication given by
$\De(a\ten b):=\sum_{(a),(b)}
(a_{(1)}\ten b_{(1)})\ten(a_{(2)}\ten b_{(2)})$.

\Lemma{20}
Let $\FSA$ and $\FSB$ be coalgebras. Let $s$ and $t$ be irreducible
co\rep s of $\FSA$ resp.\ $\FSB$ on finite dimensional vector spaces
$V$ and $W$. Define the co\rep\ $r$ of the coalgebra $\FSA\ten\FSB$ on
$V\ten W$ as the tensor product of the co\rep s $s$ and $t$:
$$
r(v\ten w)=\sum_{(v),(w)}(v_{(1)}\ten w_{(1)})\ten(v_{(2)}\ten w_{(2)}).
\eqno\eq{19}
$$
Then $r$ is irreducible.

\Proof
Let $U$ be a nonzero invariant irreducible subspace of $V\ten W$ \wrt $r$.
It will readily follow that $U=V\ten W$ if we can show that $U$ contains
a nonzero element of the form $v\ten w$.
Define the co\rep\ $\wt s$ of $\FSA$ on $V\ten W$ by
$$
\wt s(v\ten w):=\sum_{(v)}(v_{(1)}\ten w)\ten v_{(2)}.
$$
Now, because $\wt s=(\id\ten\id\ten\id\ten\ep_{\FSB})\circ r$,
we see that $U$ is invariant \wrt\ $\wt s$, so $U$ contains an
invariant subspace $U_0$ on which the restriction $s_0$ of $\wt s$ is an
irreducible co\rep\ of $\FSA$.
Choose a basis $f_1,\ldots,f_n$ for $W$. Let $P_j\colon V\ten W\to V$ be
the operator wich sends $v\ten f_j$ to
$v$ and is 0 on the other $V\ten f_k$. Let $\pi_j$ be the restriction of
$P_j$ to $U_0$. Then $\pi_j$ is an intertwining operator for $s_0$ and $s$.
By Lemma \thtag{18}(b) $\pi_j$ is bijective or 0.
Now $\pi_j$ is nonzero for some $j$; after rearranging
the $f_j$ we may assume that $\pi_1\ne0$. Then, for each $j$,
$\pi_j\circ\pi_1^{-1}\colon V\to V$ is an intertwining operator for $s$
and $s$. Thus, by Lemma \thtag{18}(c),
$\pi_j\circ\pi_1^{-1}=\la_j I$ for some complex $\la$. We conclude that
$u\in U_0$ can be written as
$u=\sum_{j=1}^n \pi_j(u)\ten f_j=
\pi_1(u)\ten\bigl(\sum_{j=1}^n\la_jf_j\bigr)$.\hhalmos

\Def{21}
Let $(t_{ij})_{i,j=1,\ldots,n}$ be a matrix corepresentation of a Hopf
algebra $\FSA$. Then, because of \eqtag{11} and \eqtag{13},
$(S(t_{ji}))_{i,j=1,\ldots,n}$ is also a matrix co\rep\ of $\FSA$.
We denote it by $(t'_{ij})$ and we call it the
{\sl contragredient co\rep} of $t$.

\bPP
Note that, if the matrix co\rep s $t$ and $s$ are equivalent with
intertwining operator given by the matrix $A$ then $t'$ and $s'$ are
equivalent with intertwining operator given by the matrix ${}^tA^{-1}$.
If $S$ is invertible (for instance if $\FSA$ is a Hopf $*$-algebra)
and if $t$ is a matrix co\rep\ then $t$ is irreducible iff
$t'$ is irreducible. (For the proof, suppose that one of both co\rep s
is not irreducible and bring it, by equivalence, in suitable block
matrix form.)

Let $\FSA$ be a Hopf algebra.
Because of \eqtag{2}, the mapping $\De\colon\FSA\to\FSA\ten\FSA$
is a co\rep\ of $\FSA$ on $\FSA$, the so-called {\sl right regular
co\rep}. Similarly, the mapping $\tau\circ(S\ten\id)\circ\De
\colon\FSA\to\FSA\ten\FSA$
defines the {\sl left regular co\rep} on $\FSA$. This is indeed a co\rep:
$$
(\tau\circ(S\ten\id)\circ\De)(a)=\sum_{(a)}a_{(2)}\ten S(a_{(1)}),
\quad a\in\FSA.
$$
Then both
$(\tau\circ(S\ten\id)\circ\De)\ten\id$
and
$\id\ten\De$ applied to the above \RHS\ yield
$$
\sum_{(a)} a_{(3)}\ten S(a_{(2)})\ten S(a_{(1)}).
$$

\Prop{22}
Let $\FSA$ be a Hopf algebra with invertible antipode.
Let $\{t^\al\}_{\al\in I}$ ($I$ some index set) be a collection
of mutually inequivalent irreducible matrix co\rep s of $\FSA$.
Then the set of all matrix elements $t_{ij}^\al$ is a set of linearly
independent elements.

\Proof
Consider first a single irreducible $d$-dimensional matrix co\rep\ $t$ of
$\FSA$. We have
$$
\eqalignno{
\De(t_{ij})&=\sum_k t_{ik}\ten t_{kj},&\eq{20}
\cr (\tau\circ(S\ten\id)\circ\De)(t_{ij})&=
\sum_k t_{kj}\ten {t'}_{ki}.&\eq{21}
\cr}
$$
Let $\wt\De$ be the co\rep\ of $\FSA\ten\FSA$ on $\FSA$ which is
defined by
$$
\wt\De:=((\tau\circ(S\ten\id)\circ\De)\ten\id)\circ\De\colon
a\mapsto\sum_{(a)} S(a_{(2)})\ten a_{(1)}\ten a_{(3)}\colon
\FSA\to\FSA\ten\FSA\ten\FSA.
$$
Then, by \eqtag{20} and \eqtag{21}, we
have
$$
\wt\De(t_{ij})=\sum_{k,l}t_{kl}\ten {t'}_{ki}\ten t_{lj}.
$$
Thus $\Span\{t_{ij}\}$ is an invariant subspace of $\FSA$ \wrt\ $\wt\De$.
Denote the restriction of the co\rep\ $\wt\De$ to this subspace by
$\wh t$.

Let $e_1,\ldots,e_d$ be the standard basis of $\CC^d$. We can consider
both $t$ and $t'$ as co\rep s of $\FSA$ on $\CC^d$, cf.\ \eqtag{14}. Let
$\wt t$ be the co\rep\ of $\FSA\ten\FSA$ on $\CC^d\ten\CC^d$ which is the
tensor product of the co\rep s $t'$ and $t$ of $\FSA$ on $\CC^d$.  By
Lemma \thtag{20} this co\rep\ is irreducible. Then
$$
\wt t(e_i\ten e_j)=\sum_{k,l}(e_k\ten e_l)\ten{t'}_{ki}\ten t_{lj}.
$$
Thus the surjective linear mapping
$A\colon\CC^d\ten\CC^d\to\Span\{t_{ij}\}$ defined by
$A(e_i\ten e_j):=t_{ij}$ is an intertwining operator for the co\rep s
$\wt t$ and $\wh t$. Then $A^{-1}(0)$ is an invariant subspace of
$\CC^d\ten\CC^d$ \wrt\ $\wt t$, certainly a proper subspace,
since otherwise $t_{11}=0$, while $\ep(t_{11})=1$.
Because of
the irreducibility of $\wt t$ we then have $A^{-1}(0)=\{0\}$. Thus $A$ is
bijective. Hence the $t_{ij}$ are linearly independent.

Next consider $m$
mutually inequivalent irreducible matrix co\rep s $t^l$ ($l=1,\ldots,m$)
of $\FSA$, where
$t^l=(t_{ij}^l)_{i,j=1,\ldots,d_l}$.
Denote the linear span of all elements $t_{ij}^l$ by $V$.
Then, with the above notation, $V$ is an invariant subspace of $\FSA$
\wrt\ the co\rep\ $\wt\De$ of $\FSA\ten\FSA$ on $\FSA$.
Let $W:=\oplus_{l=1}^m W_l$, where $W_l:=\CC^{d_l}\ten\CC^{d_l}$.
On $W$ we have the co\rep\ of $\FSA\ten\FSA$ which is the direct sum
of the tensor product co\rep s of $(t^l)'$ with $t^l$ and the $W_l$ are
invariant irreducible subspaces of $W$. The irreducible co\rep s of
$\FSA\ten\FSA$ on the $W_l$ are mutually inequivalent.

Similarly as above we have a surjective intertwining operator
$A\colon W\to V$ \wrt\ these two co\rep s of $\FSA\ten\FSA$.
Then $A^{-1}(0)$ is an invariant subspace of $W$.
Suppose that $A^{-1}(0)\ne\{0\}$. Then we can take an invariant
irreducible subspace $U$ of $A^{-1}(0)$.
By Lemma \thtag{19}, $U=W_l$ for some $l$. This implies that, for this
$l$, $t_{ij}^l=0$, which contradicts that $\ep(t_{11}^l)=1$.\hhalmos

\Lemma{22a}
Let $\FSA$ be a Hopf algebra with invertible antipode.
Let $s$ and $t$ be irreducible matrix co\rep s of $\FSA$.
Then $s$ and $t$ are equivalent iff $\Span\{s_{ij}\}=
\Span\{t_{ij}\}$.

\Proof
Clearly, if $s$ and $t$ are equivalent then the two spans coincide.
Conversely, let $V:=\Span\{s_{ij}\}=\Span\{t_{ij}\}$. Then $V$ is
an invariant subspace of $\FSA$ \wrt the co\rep\ $\De$ of $\FSA$ on
$\FSA$. By Proposition \thtag{22} and formula \eqtag{20} the restriction
of $\De$ to $V$ is a co\rep\ of $\FSA$ which is a direct sum of
$d$ copies of $s$ and also a direct sum of $d$ copies of $t$.
By Lemma \thtag{19} we conclude that $s$ is equivalent to $t$.\hhalmos

\Def{23}
Let $s$ and $t$ be co\rep s of a Hopf algebra $\FSA$ on vector spaces $V$
and $W$, respectively. Then the {\sl tensor product} $s\ten t$ of $s$
and $t$ is defined as the co\rep\ of $\FSA$ on $V\ten W$ given by the
mapping
$(\id\ten\id\ten m)\circ(\id\ten\tau\ten\id)\circ (s\ten t)\colon
V\ten W\to (V\ten W)\ten\FSA$, i.e.,
$$
(s\ten t)(v\ten w)=\sum_{(v),(w)} v_{(1)}\ten w_{(1)}\ten v_{(2)}w_{(2)}.
$$
If $s$ and $t$ are matrix co\rep s then $s\ten t$ can also be considered
as a matrix co\rep:
$$
(s\ten t)_{ij,kl}:=s_{ik}\,t_{jl}.
$$
Note that this tensor product is associative, but not commutative.
The present tensor product should not be confused with the
co\rep\ of $\FSA\ten\FSA$ obtained as a tensor product of $s$ and $t$.

\Def{24}
We call a co\rep\ of a Hopf $*$-algebra {\sl unitarizable} if it is
equivalent to a unitary co\rep. In particular, a matrix co\rep\ $t$ of
$\FSA$ is unitarizable iff there is an invertible complex square matrix
$A$ \st\
$s:=AtA^{-1}$ satisfies $S(s_{ij})=s_{ji}^*$.

\bPP
Note that tensor products of unitary (resp.\ unitarizable) matrix co\rep s
are again unitary (resp.\ unitarizable).

\Sec2 {Compact quantum groups}
This section gives a new approach to the theory of compact matrix
quantum groups as developed by Woronowicz \ref{Wor87},
\ref{Wor88}, \ref{Wor91}.
While Woronowicz's definition already involves $C^*$-algebras and
his further development of the theory also heavily uses $C^*$-algebra
theory,
the present approach only uses Hopf $*$-algebras, and mentions the
connection
with $C^*$-algebras at a much later stage, more as a side remark.
This new approach was developed in cooperation with M. S. Dijkhuizen
(CWI, Am\-ster\-dam). It will also be described in his forthcoming Ph.D.\
thesis \ref{Dij94}.
The relationship with other work on compact quantum groups will be
discussed
at the end of this section.
In particular, we acknowledge that Effros \& Ruan \ref{EfRu93}
introduced CQG algebras (named differently by them) earlier.
However, they developed their theory in a very different way.

\Subsec1 {CQG algebras and CMQG algebras}
Let $G$ be a compact group.
Let $\Pi(G)$ be the linear span  of all matrix elements of
irreducible unitary (hence finite dimensional) \rep s of $G$. Then $\Pi(G)$
is a commutative Hopf $*$-algebra with Hopf $*$-algebra operations
as described in
Examples \thtag{5} and \thtag{62}. We speak of a
{\sl compact matrix group} if, for some $n$, $G$ is (isomorphic to) a
closed subgroup of the group $U(n)$ of $n\times n$ unitary marices.
For a compact group $G$ there are the following equivalent
statements: G is a compact matrix group iff G is a compact Lie group
iff $\Pi(G)$ is finitely generated.
This suggests the following definitions.

\Def{25}
A Hopf $*$-algebra $\FSA$ is said to be associated with a {\sl
compact quantum group} if it is the linear span of the matrix elements
of the unitary (finite -dimensional) matrix co\rep s of $\FSA$.
Then $\FSA$ is called a {\sl CQG algebra}.

\bPP
Because of Proposition \thtag{17}(b), a CQG algebra $\FSA$ is also the
linear span of the matrix elements of its irreducible unitary matrix
co\rep s.
Denote by $\Ahat$ the collection of all equivalence classes of
irreducible unitary matrix co\rep s of $\FSA$. Choose,
for each $\al\in\Ahat$, a unitary matrix co\rep\
$(t_{ij}^\al)_{i,j=1,\ldots,d_\al}$ belonging to class $\al$.
Then, by Proposition \thtag{22}, the set of all $t_{ij}^\al$ forms a
basis of $\FSA$. Write
$$
\FSA_\al:=\Span\{t_{ij}^\al\}_{i,j=1,\ldots,d_\al},\quad\al\in\wh\FSA.
$$
Then $\FSA$ is the direct sum of the $\FSA_\al$.
Let 1 be the element of $\Ahat$ for which $t^1$ is the one-dimensional
matrix co\rep\ $(1)$.
For $\al\in\Ahat$ let $\al'$ be the element of $\Ahat$ \st\
$(t^\al)'$ is equivalent to $t^{\al'}$.

\Prop{26}
Let $\FSA$ be a CQG algebra and let the $t^\al$ be as above.
\item{(a)}
Each irreducible matrix co\rep\ of $\FSA$ is
equivalent to some $t^\al$ ($\al\in\Ahat$).
\item{(b)}
Each irreducible matrix co\rep\ of $\FSA$ is unitarizable.
\item{(c)}
If $s$ is a unitarizable matrix co\rep\ of $\FSA$ then so is $s'$.

\Proof
(b) immediately follows from (a). For (c) observe that a unitarizable $s$
is a direct sum of irreducible co\rep s. Hence $s'$ is also a direct sum
of irreducible co\rep s. Then apply (b).
Thus we have to prove (a).
Let $s$ be an irreducible matrix co\rep\ of $\FSA$.
The spaces $\FSA_\al$ are invariant irreducible subspaces of $\FSA$
\wrt the co\rep\ of $\FSA\ten\FSA$ on $\FSA$ which is the tensor product
of the left regular and the right regular co\rep\ (cf.\ the proof of
Proposition \thtag{22}), and so is $V:=\Span\{s_{ij}\}$.
Moreover, the co\rep s obtained by restriction to the various $\FSA_\al$
are mutually inequivalent. By Lemma \thtag{19} $V=\FSA_\al$ for some
$\al\in\Ahat$. By Lemma \thtag{22a} we see that $s$ is equivalent to
$t^\al$.\hhalmos

\Def{27}
A Hopf $*$-algebra $\FSA$ is said to be associated with a {\sl compact
matrix quantum group} if it is a finitely generated CQG algebra.
Then $\FSA$ is called a {\sl CMQG algebra}.

\Prop{28}
For a Hopf $*$-algebra the following conditions are equivalent.
\item{(i)}
$\FSA$ is a CMQG algebra.
\item{(ii)}
There is a unitary matrix co\rep\ $v$ of $\FSA$ \st\ $\FSA$ is generated
as an algebra by the matrix coefficients of $v$.
\item{(iii)}
There is a matrix co\rep\ $u$ of $\FSA$ \st\ both $u$ and $u'$
are unitarizable and $\FSA$ as an algebra is generated
by the matrix elements of $u$ and $u'$.

\Proof
\LP
(i)$\Rightarrow$(ii)$\Rightarrow$(iii):\quad
Assume (i). Each of the
generators is a linear combination of matrix
elements of unitary matrix co\rep s of $\FSA$. Hence, by taking direct
sums, there is a unitary matrix co\rep\ $u$ of $\FSA$ of which the matrix
elements generate $\FSA$. Then $u'$ is unitarizable by Proposition
\thtag{26}.
\sLP
(iii)$\Rightarrow$(ii)$\Rightarrow$(i):\quad
Assume (iii). Let $v$ be the direct sum of two unitary matrix co\rep s
which are equivalent to
$u$ respectively $u'$. Then $\FSA$ is generated by the matrix elements
of $v$.
Each product of matrix elements of $v$ is
a matrix element of some multiple tensor product of $v$.
Such tensor products are again unitary co\rep s.
Thus $\FSA$ is a finitely generated CQG algebra.\hhalmos

\Subsec2 {The Haar functional}
Define on a CQG algebra $\FSA$ the {\sl Haar functional} $h$ as the linear
mapping $h\colon\FSA\to\CC$ \st
$$
h(a):=\cases{
0&if $a\in\FSA_\al$, $\al\ne1$,
\cr
1&if $a=1$,
\cr}
\eqno\eq{22}
$$
The functional $h$ satisfies
$$
(h\ten\id)(\De(a))=h(a)\,1=(\id\ten h)(\De(a)),\quad a\in\FSA.
\eqno\eq{23}
$$
Indeed, if $a$ equals some basis element $t_{ij}^\al$
of $\FSA$ then \eqtag{23} takes the form
$$
\sum_k h(t_{ik}^\al)\,t_{kj}^\al=h(t_{ij}^\al)\,1=
\sum_k h(t_{kj}^\al)\,t_{ik}^\al,
\eqno\eq{24}
$$
and these identities are indeed implied by \eqtag{22}.

On the other hand, if $h$ would be any linear functional on
$\FSA$ satisfying the first (or the second) identity in \eqtag{23}
then we obtain from \eqtag{24} by linear independence of the $t_{ij}^\al$
that $h(t_{ij}^\al)=0$ for $\al\ne1$. Thus, $h$ is determined up to a
constant factor by each of the identities in \eqtag{23} and $h$ can next
be normalized by
$$
h(1)=1.
\eqno\eq{24a}
$$

Since $S(\FSA_\al)=\FSA_{\al'}$, \eqtag{22} implies that
$$
h(S(a))=h(a),\quad a\in\FSA.
\eqno\eq{25}
$$

If $G$ is a compact group then the normalized Haar measure, as a linear
functional on $\Pi(G)$, can be interpreted as a Haar functional.
Indeed,
$$
h(a):=\int_G a(x)\,dx,\quad a\in\Pi(G),
$$
and
$$
((h\ten\id)(\De(a)))(y)=\int_G a(xy)\,dx=\int_G a(x)\,dx=h(a)\,1(y),
$$
and similarly for the other identity.
In this case we also have that $h(aa^*)=\int_G |a(x)|^2\,dx>0$
for $a\ne0$. We want to prove this positivity result also for $h$ in the
case of a CQG algebra.

The following Lemma will be crucial.
For the moment we will assume that $\FSA$ is a Hopf algebra with
invertible antipode (not necessarily a Hopf $*$-algebra)
and that
$h$ is a linear functional on $\FSA$ satisfying \eqtag{23}, \eqtag{24a}
and \eqtag{25}.

\Lemma{30}
Let $r$ and $s$ be matrix co\rep s of $\FSA$.
Then
$$
\eqalignno{
\sum_l h(s_{ij}\,S(r_{kl}))\,r_{lm}&=
\sum_l s_{il} \,h(s_{lj}\,S(r_{km})),&\eq{26}
\cr
\sum_l h(S(r_{ij})\,s_{kl})\,s_{lm}&=
\sum_l r_{il}\,h(S(r_{lj})\,s_{km}).&\eq{27}
\cr}
$$
With the notation
$$
A_{il}^{(j,k)}:=h(s_{ij}\,S(r_{kl})),\quad
B_{il}^{(j,k)}:=h(S(r_{ij})\,s_{kl}),\eqno\eq{28}
$$
the identities \eqtag{26} and \eqtag{27} can be rewritten as
$$
A^{(j,k)}\,r=s\,A^{(j,k)},\quad
B^{(j,k)}\,s=r\,B^{(j,k)}.\eqno\eq{29}
$$
Thus $A^{(j,k)}$ is an intertwining operator for $r$ and $s$ and
$B^{(j,k)}$ is an intertwining operator for $s$ and $r$.

\Proof
For the proof of \eqtag{26} write
$$
\displaylines{
\quad h(s_{ij}\,S(r_{kl}))\,1=
(\id\ten h)(\De(s_{ij}\,S(r_{kl})))\hfill
\cr
\hfill=\sum_{p,n}(\id\ten h)((s_{ip}\ten s_{pj})\,(S(r_{nl})\ten S(r_{kn}))
=\sum_{p,n} h(s_{pj}\,S(r_{kn}))\,s_{ip}\,S(r_{nl}).\quad
\cr}
$$
Substitute this equality in the \LHS\ of \eqtag{26} and next use that
$\sum_l S(r_{nl})\,r_{lm}\allowbreak=\de_{nm}\,1$. This settles \eqtag{26}.
Similarly, \eqtag{27} is obtained from a substitution of
$$
h(S(r_{lj})\,s_{km})\,1=(h\ten\id)(\De(S(r_{lj})\,s_{km}))
$$
in the \RHS\ of \eqtag{27}.\hhalmos

\bPP
Let us assume that $r$ and $s$ are irreducible matrix co\rep s of
$\FSA$. We will consider three special cases of Lemma
\thtag{30}: (i) $r$ and $s$ are inequivalent, (ii) $s$=$r$,
(iii) $s=r''$ (i.e., $s_{ij}=S^2(r_{ij})$).
For each case we apply one of the quantum Schur lemmas (Lemma
\thtag{18}(b),(c)).
\LP{\sl Case (i)}\quad
If $r$ and $s$ are inequivalent then \eqtag{29} yields
that $A_{il}^{(j,k)}=0$ and $B_{il}^{(j,k)}=0$.
Hence, by \eqtag{28},
$$
h(s_{ij}\,S(r_{kl}))=0,\quad
h(S(r_{ij})\,s_{kl})=0.
$$
\LP{\sl Case (ii)}\quad
If $s=r$ then \eqtag{29} yields that there are complex constants
$\al_{jk}$ and $\be_{jk}$ \st\
$A^{(j,k)}=\al_{jk}\,I$ and $B_{il}^{(j,k)}=\be_{jk}\,I$.
Hence, by \eqtag{28}
$$
\eqalignno{
A_{il}^{(j,k)}=h(r_{ij}\,S(r_{kl}))&=\al_{jk}\,\de_{il},&\eq{29a}
\cr
B_{il}^{(j,k)}=h(S(r_{ij})\,r_{kl})&=\be_{jk}\,\de_{il}.&\eq{29b}
\cr}
$$
Moreover, if we fix $i=l$ and sum over $j=k$ then we obtain
$$
\sum_j \al_{jj}=1=\sum_j \be_{jj}.
\eqno\eq{29c}
$$
\LP{\sl Case (iii)}\quad
If $s=r''$ then \eqtag{26} and \eqtag{27} together with \eqtag{25}
yield
$$
\eqalignno{
\sum_l h(r_{kl}\,S(r_{ij}))\,r_{lm}&=
\sum_l S^2(r_{il})\,h(r_{km}\,S(r_{lj})),
\cr
\sum_l h(S(r_{kl})\,r_{ij})\,S^2(r_{lm})&=
\sum_l r_{il}\,h(S(r_{km})\,r_{lj}).
\cr}
$$
Put
$$
\eqalignno{
&\wt A_{il}^{(j,k)}:=h(r_{kl}\,S(r_{ij}))=A_{kj}^{(l,i)}
=\al_{li}\,\de_{kj},&\eq{30}
\cr
&\wt B_{il}^{(j,k)}:=h(S(r_{kl})\,r_{ij})=B_{kj}^{(l,i)}
=\be_{li}\,\de_{kj},&\eq{31}
\cr}
$$
where we refer to \eqtag{29a}, \eqtag{29b}.
Then $\wt A^{(j,k)}\,r=r''\,\wt A^{(j,k)}$
and $\wt B^{(j,k)}\,r''=r\,B^{(j,k)}$.
Moreover, when we sum the first equality in \eqtag{30} and in
\eqtag{31} over
$i=l$ then we obtain that
$$
\tr \wt A^{(j,k)}=\de_{jk},\quad\tr \wt B^{(j,k)}=\de_{jk}.
$$
Hence, there exists a nonzero intertwining operator $F$ for $r$ and $r''$.
Thus $r$ and $r''$ are equivalent irreducible co\rep s
(since $S$ is assumed to be invertible, the irreducibility of $r$
implies the irreducibility of $r''$),
and $F$ is an
invertible operator, unique up to a constant factor and satisfying
$\tr F\ne0$ and $\tr F^{-1}\ne0$.
Hence there are complex constants $\wt\al_{jk}$ and
$\wt\be_{jk}$ \st
$$
\eqalignno{
\wt A_{il}^{(j,k)}&=\wt\al_{jk}\,F_{il}=\al_{li}\,\de_{kj},
\cr
\wt B_{il}^{(j,k)}&=\wt\be_{jk}\,(F^{-1})_{il}=\be_{li}\,\de_{kj},
\cr}
$$
where, in each line, the second equality follows from
\eqtag{30}. Combination with \eqtag{29c} yields that
$$
\wt\al_{jk}\,\tr F=\de_{kj},\quad
\wt\be_{jk}\,\tr F^{-1}=\de_{kj}.
$$
Hence
$$
\wt A_{il}^{(j,k)}=\de_{kj}\,{F_{il}\over\tr F}\,,\quad
\wt B_{il}^{(j,k)}=\de_{kj}\,{(F^{-1})_{il}\over\tr F^{-1}}\,.
\eqno\eq{32}
$$

Let us summarize the obtained results in the following Proposition.

\Prop{31}
Let $\FSA$ be a Hopf algebra with invertible antipode
and let $h$ be a linear functional on $\FSA$
satisfying \eqtag{23}, \eqtag{24a} and \eqtag{25}.
Let $r$ be an irreducible matrix co\rep\ of $\FSA$.
\item{(a)}
If $s$ is an irreducible matrix co\rep\ of $\FSA$ which is not equivalent
to $r$ then
$$
h(s_{kl}\,S(r_{ij}))=0,\quad h(S(r_{kl})\,s_{ij})=0.
\eqno\eq{32a}
$$
\item{(b)}
$r$ is equivalent to $r''$. Let $F$ be an invertible intertwining operator
for $r$ and $r''$. Then $\tr F\ne0$ and $\tr F^{-1}\ne0$ and
$$
\eqalignno{
h(r_{kl}\,S(r_{ij}))&=\de_{kj}\,{F_{il}\over\tr F}\,,&\eq{33}
\cr
h(S(r_{kl})\,r_{ij})&=\de_{kj}\,{(F^{-1})_{il}\over\tr F^{-1}}\,.&\eq{34}
\cr}
$$
\Proof
\eqtag{33} and \eqtag{34} were obtained from \eqtag{30}, \eqtag{31} and
\eqtag{32}.\hhalmos

\bPP
Now we assume that $\FSA$ is a CQG algebra.

\Prop{32}
Let $r$ be an irreducible unitary matrix co\rep\ of the CQG algebra
$\FSA$. Let the matrix $F$ be an invertible
intertwining operator for $r$ and $r''$ as in Proposition \thtag{31}.
Then $F$ is a constant multiple of a positive definite matrix. It can be
uniquely normalized
\st\ $\tr F=\tr F^{-1}>0$.

\Proof
By Proposition \thtag{26}(c) there is a unitary matrix co\rep\ $s$
\st\ $r'$ is equivalent to $s$. Thus there is an invertible complex matrix
$A$ \st\ $sA=Ar'$. Let $\wb A$ be the matrix for which
$(\wb A)_{ij}:=\wb{A_{ij}}$ and let ${}^tA$ be the transpose of $A$.
Since ${s'}_{ij}=(s_{ij})^*=S(s_{ji})$ and
${r'}_{ij}=(r_{ij})^*$, ${r''}_{ij}=S({r'}_{ji})$, we conclude from
$sA=Ar'$ that $s'\,\wb A=\wb A\,r$ and
${}^tA\,s'=r''\,{}^tA$. Hence
${}^tA\,\wb A\,r=r''\,{}^tA\,\wb A$.
Thus $F=\const {}^tA\,\wb A$ and ${}^tA\,\wb A$ is positive
definite.\hhalmos

\Theor{33}
The Haar functional on a CQG algebra satisfies
$$
h(aa^*)>0\quad\hbox{if $a\ne0$,}
$$
\Proof
Propositions \thtag{31} and \thtag{32} imply that
$$
h(t_{kl}^\al\,(t_{ij}^\be)^*)=\de_{\al\be}\,\de_{ki}\,G_{lj}^\al,\quad
\al,\be\in\wh\FSA,
$$
for certain positive definite matrices $G^\al$.
Let $a:=\sum_{\al,k,l}c_{kl}^\al\,t_{kl}^\al$ be an arbitrary element of
$\FSA$. Then
$$
h(aa^*)=\sum_{\al,k}\,\sum_{l,j}c_{kl}^\al\,\wb{c_{kj}^\al}\,G_{lj}^\al
\ge0.
$$
If $h(aa^*)=0$ then
$\sum_{l,j}c_{kl}^\al\,\wb{c_{kj}^\al}\,G_{lj}^\al=0$ for all $\al$ and
$k$. By positive definiteness of the $G^\al$ this implies that all
coefficients $c_{kl}^\al$ are 0, i.e., $a=0$.\hhalmos

\Cor{34}
Every finite dimensional co\rep\ of a CQG algebra $\FSA$ is
unitarizable, and hence decomposable as a direct sum of irreducible
finite dimensional co\rep s.
\Proof
Let $r$ be a co\rep\ of $\FSA$ on a vector space
$V$ with some inner product $(\,.\,,\,.\,)$.
Define a new sesquilinear form on $V$ by
$$
\lan v,w\ran:=\sum_{(v),(w)}(v_{(1)},w_{(1)})\,h(w_{(2)}^*\,v_{(2)}).
$$
Then $r$ is unitary in the sense of \eqtag{16}
\wrt this sesquilinear form. Indeed,
$$
\eqalignno{
\sum_{(v),(w)}\lan v_{(1)},w_{(1)}\ran\,w_{(2)}^*\,v_{(2)}&=
\sum_{(v),(w)}(v_{(1)},w_{(1)})\,h(w_{(2)}^*\,v_{(2)})\,w_{(3)}^*\,v_{(3)}
\cr
&=\sum_{(v),(w)}(v_{(1)},w_{(1)})\,h(w_{(2)}^*\,v_{(2)})\,1=
\lan v,w\ran\,1.
\cr}
$$
It is left to show that $\lan\,.\,,\,.\,\ran$ is positive definite.
Let $e_1,\ldots,e_n$ be an orthonormal basis of $V$ \wrt the inner
product $(\,.\,,\,.\,)$ and write $r$ as a matrix co\rep\ \wrt this
basis. Then
$$
\lan e_i,e_j\ran=\sum_{k,l}(e_k,e_l)\,h(r_{lj}^*\,r_{ki})=
\sum_k h(r_{kj}^*\,r_{ki}).
$$
Hence
$$
\textstyle\lan\sum_i c_ie_i,\sum_j c_j e_j\ran=
\sum_k h\bigl(\bigl(\sum_j c_j r_{kj}\bigr)^*\,
\bigl(\sum_i c_i r_{ki}\bigr)\bigr)\ge0,
$$
and if the \LHS\ equals 0 then $\sum_i c_i\,r_{ki}=0$ for all $k$,
hence $c_k=\ep\bigl(\sum_i c_i\,r_{ki}\bigr)\allowbreak=0$ for al $k$.\hhalmos

\Subsec3 {The $C^*$-completion of a CQG algebra}
Recall that a {\sl $C^*$-algebra} is a Banach algebra $A$ which is also a
$*$-algebra \st\ $||aa^*||=||a||^2$ for all $a\in A$.
In particular, the space $\FSL(\FSH)$ of all bounded linear operators
on a Hilbert space $\FSH$ forms a $C^*$-algebra with identity.
By a {\sl $*$-\rep} of a $*$-algebra $\FSA$ on a Hilbert space $\FSH$ we
mean a $*$-algebra homomorphism $\pi\colon\FSA\to\FSL(\FSH)$.

On a compact group $G$ the $*$-algebra $\Pi(G)$ can be
equipped with the sup norm. The space $C(G)$ of all continuous functions
on $G$ is the completion of $\Pi(G)$ \wrt this norm and $C(G)$ is a
commutative
$C^*$-algebra with identity. Both for the $*$-algebra $\Pi(G)$ and for the
$C^*$-algebra $C(G)$ the irreducible $*$-\rep s on a Hilbert space are
precisely the one-dimensional $*$-homomorphisms $a\mapsto a(x)$
($x\in G$).

This suggests the following strategy for the construction of an
analogue of $C(G)$ for a CQG algebra $\FSA$:
\item{1)}
Define a seminorm
$$
||a||:=\sup_\pi ||\pi(a)||
\eqno\eq{35}
$$
on $\FSA$, where the supremum runs over all $*$-\rep s $\pi$
of the $*$-algebra $\FSA$.
\item{2)}
Show that this seminorm is a norm.
\item{3)}
Let $A$ be the completion of $\FSA$ \wrt this norm.

\mPP
First observe that $||a||$ defined by \eqtag{35} is finite for all
$a\in\FSA$. Indeed, if $a$ is expanded in terms of the basis of $\FSA$
by the finite sum
$$
a=\sum_{\al,i,j}c_{ij}^\al\,t_{ij}^\al
\eqno\eq{36}
$$
then
$||a||\le\sum_{\al,i,j}|c_{ij}^\al|\,||t_{ij}^\al||$.
Now use the following lemma.

\Lemma{35}
Let $r$ be a unitary matrix co\rep\ of a CQG algebra $\FSA$.
Let $V$ be a complex vector space with hermitian inner product and
let $|v|:=(v,v)^{\half}$ define the norm on $V$.
Let $\pi$ be an algebra homomorphism from $\FSA$ to the algebra
of all linear operators on $V$ \st\
$(\pi(a)v,w)=(v,\pi(a^*)w)$ for all $a\in\FSA$ and all $v,w\in V$.
Then $\pi(r_{ij})$ is a bounded linear operator on $V$
of norm $\le1$ for all $i,j$. In particular, $\pi$ uniquely extends to a
$*$-algebra \rep\ of $\FSA$ on the Hilbert space completion of $V$.

\Proof
Since $r$ is unitary, we have $\sum_k r_{kj}^*\,r_{kj}=1$.
Hence, for all $v\in V$,
$$
||v||^2=(v,v)=\sum_k (\pi( r_{kj}^*\,r_{kj})\,v,v)
=\sum_k (\pi(r_{kj})\,v,\pi(r_{kj})\,v)\ge||\pi(r_{ij})\,v||^2.
\eqno\halmos
$$

\bPP
One consequence of this Lemma is that, for $a\in\FSA$ with expansion
\eqtag{36}, all norms $||\pi(a)||$, and hence $||a||$, are bounded by
$\sum_{\al,i,j}|c_{ij}^\al|$.

Another consequence concerns the left regular \rep\ of $\FSA$ on $\FSA$,
which is defined by $\la(a)\,b:=ab$. This is an algebra \rep, which is
{\sl faithful}, i.e., if $\la(a)=0$ then $a=0$.
Now equip $\FSA$ with the inner product $(a,b):=h(b^*a)$.
(By the properties of $h$ this is indeed an inner product.)
Clearly, $(\la(a)\,b,c)=(b,\la(a^*)\,c)$. Since each $a\in\FSA$ is a
linear combination of matrix elements of unitary co\rep s,
it follows by Lemma \thtag{35} that each $\la(a)$ is a bounded linear
operator on the inner product space $\FSA$. Hence $\la$ extends to a
$*$-\rep\ of $\FSA$ on the Hilbert space completion of this inner
product space.
(I thank P. Podle{\'s} for this observation.)
By faithfulness it follows that the seminorm \eqtag{35}
is a norm.

Because $||\pi(ab)||\le||\pi(a)||\,||\pi(b)||$ and
$||\pi(a)||^2=||\pi(aa^*)||$ for each $*$-rep\-re\-sen\-ta\-tion
$\pi$ of $\FSA$,
it follows that the norm \eqtag{35} is a $C^*$-norm, i.e.,
$||ab||\le||a||\,||b||$ and
$||aa^*||=||a||^2$. Thus the norm completion $A$ of $\FSA$ is a
well-defined $C^*$-algebra with identity.

Since, for each element $a$ of a $C^*$-algebra $A$, there exists an
irreducible $*$-\rep\ $\pi$ of $A$ for which $||\pi(a)||=||a||$
(cf.\ for instance Arveson \reff{Arv76}{Corollary to Theorem 1.7.2}),
the norm definition \eqtag{35} does not change when we take the supremum
there only over the irreducible $*$-\rep s of $\FSA$.

The counit and comultiplication on $\FSA$  have unique continuous
extensions to $A$. For the counit this is obvious, since $\ep$ is
a one-dimensional $*$-\rep\ of $\FSA$, so $|\ep(a)|\le||a||$.

For the $C^*$-extension of $\De$ we need a suitable $C^*$-norm on the
algebraic tensor product
$A\ten A$. We choose the {\sl injective cross norm}, i.e., for
$a\in A\ten A$, the norm $||a||$ is defined as the supremum of
the numbers $||(\pi_1\ten\pi_2)(a)||$, where $\pi_1$ and $\pi_2$ run
over the $*$-\rep s of the $C^*$-algebra $A$.
Then $\De$ continuously extends to a mapping from $A$ to the completion of
$A\ten A$ \wrt this norm, since
$a\mapsto(\pi_1\ten\pi_2)(\De(a))$ is a $*$-\rep\ of $\FSA$ for any two
$*$-\rep s $\pi_1$ and $\pi_2$ of $\FSA$, so
$||(\pi_1\ten\pi_2)(\De(a))||\le||a||$.
(The $*$-\rep\  $a\mapsto(\pi_1\ten\pi_2)(\De(a))$ of $\FSA$ is called the
{\sl tensor product} $\pi_1\ten\pi_2$ of the $*$-\rep s $\pi_1$ and
$\pi_2$, cf.\ Remark \thtag{6a}.)

\Subsec4 {A class of multiplicative linear functionals}
Let $\FSA$ be a CQG algebra. According to Proposition \thtag{32}
we can choose for each $\al\in\wh\FSA$ a positive definite complex matrix
$F_\al$ which is an intertwining operator for the matrix co\rep s
$t^\al$ and $(t^\al)''$ and which is uniquely normalized \st\
$\tr F_\al=\tr (F_\al)^{-1}$.
We can now rewrite \eqtag{32a}, \eqtag{33} and \eqtag{34} as
$$
\eqalignno{ h(t_{kl}^\be\,(t_{ji}^\al)^*)&=
\de_{\al\be}\,\de_{kj}\,{(F_\al)_{il}\over\tr F_\al}\,,&\eq{47}
\cr h((t_{lk}^\be)^*\,t_{ij}^\al)&=
\de_{\al\be}\,\de_{kj}\,{(F_\al^{-1})_{il}\over\tr F_\al^{-1}}\,.&\eq{48}
\cr}
$$
These can be considered as the analogues for CQG algebras of
Schur's orthogonality relations.
Woronowicz, for the case of his compact matrix quantum groups,
gave these quantum Schur orthogonality relations in
\reff{Wor87}{(5.14), (5.15)}.

The remainder of this subsection is an account of
Theorem 5.6 in Woronowicz \ref{Wor87}.
It describes the so-called modular properties of the Haar functional $h$.
Write $F_\al$ as
$$
F_\al=U\,\pmatrix{\la_1&&0\cr&\ddots&\cr0&&\la_{d_\al}}\,U^{-1},
$$
where $U$ is a unitary matrix and $\la_1,\ldots,\la_{d_\al}>0$.
Now define complex powers $F_\al^z$ ($z\in\CC$) of $F_\al$ by
$$
F_\al^z:=
U\,\pmatrix{\la_1^z&&0\cr&\ddots&\cr0&&\la_{d_\al}^z}\,U^{-1}.
$$
Then the matrix elements $(F_\al^z)_{ij}$ are entire analytic functions
of $z$ and there are constants $M>0$ and $\mu\in\RR$ \st\
$$
|(F_\al^z)_{ij}|\le M\,e^{\mu\Re z}\quad\hbox{for all $z\in\CC$.}
$$
Note also that $F_\al^z\,F_\al^{z'}=F_\al^{z+z'}$.

\Remark{36}
If $f$ is an entire analytic function satisfying the estimate
$|f(z)|\le M\,e^{\mu\Re z}$ for some $M>0$ and $\mu\in\RR$ then
we can find a real constant $c$ \st\ the function $g(z):=e^{cz}\,f(z)$
satisfies the estimate
$$
\exists M>0\quad\exists\mu\in(0,\pi)\qquad
\Re z\ge0\Longrightarrow|g(z)|\le M\,e^{\mu|z|}.
\eqno\eq{37}
$$
{\sl Carlson's theorem} (cf.\ for instance Titchmarsh
\reff{Tit39}{\S5.81}
states that, if a holomorphic function $g$ on
$\{z\mid \Re z\ge0\}$ satisfies \eqtag{37} and if moreover
$g(1)=g(2)=\cdots=0$, then $g$ will vanish identically.
Hence, if the above function $f$ will vanish on the set $\{1,2,\ldots\}$
then it will also vanish identically.

\bPP
Now define, for each $z\in\CC$, the element $f_z\in\FSA^*$ by
specifying it on our linear basis of~$\FSA$:
$$
f_z(t_{ij}^\al):=(F_\al^z)_{ij}.
$$
Note that this definition is independent of the choice of the
matrix co\rep\ $t^\al$ in the class $\al\in\wb\FSA$.
The following properties of the $f_z$ quickly follow from the definition
of~$f_z$.

\Prop{37}
\item{(a)}
$f_1(t_{ij}^\al)=(F_\al)_{ij}$,\quad
$f_{-1}(t_{ij}^\al)=((F_\al)^{-1})_{ij}$.
\item{(b)}
$f_0=\ep$.
\item{(c)}
$f_z(1)=1$.
\item{(d)}
$f_z\,f_{z'}=f_{z+z'}$.
\item{(e)}
For each $a\in\FSA$ the function $z\mapsto f_z(a)$ is an entire analytic
function and there are constants $M>0$ and $\mu\in\RR$ \st\
$|f_z(a)|\le M\,e^{\mu\Re z}$.

\bPP
For some further properties we will give the proof.

\Prop{38}
$S^2(a)=f_{-1}.a.f_1$.
\Proof
It follows from Proposition \thtag{32} that
$$
\eqalignno{
S^2(t_{ij}^\al)&=\sum_{k,l}(F_\al)_{ik}\,t_{kl}^\al\,(F_\al^{-1})_{lj}
\cr
&=(f_1\ten\id\ten f_{-1})((\De\ten\id)\,\De(t_{ij}^\al))
\cr
&=
f_{-1}.t_{ij}^\al.f_1.&\halmos
\cr}
$$

\Prop{39}
If $a,b\in\FSA$ then
$$
h(ab)=h(b\,(f_1.a.f_1)).
\eqno\eq{45}
$$
\Proof
Let $\al,\be\in\wh\FSA$. It is enough to verify \eqtag{45} for
$a:=t_{kl}^\be$ and $b:=(t_{ij}^\al)^*$.
We apply \eqtag{47} and \eqtag{48}.
If $\al\ne\be$ then both sides of \eqtag{45} are 0.
If $\al=\be$ then
$$
\eqalignno{
h((t_{ij}^\al)^*\,(f_1.t_{kl}^\al.f_1))&=
\sum_{m,n}(F_\al)_{km}\,(F_\al)_{nl}\,h((t_{ij}^\al)^*\,t_{mn}^\al)
\cr
&=\sum_{m,n}(F_\al)_{km}\,(F_\al)_{nl}\,{\de_{jn}\,
(F_\al^{-1})_{mi}\over\tr F_\al^{-1}}
\cr
&={\de_{ik}\,(F_\al)_{jl}\over\tr F_\al^{-1}}
={\de_{ik}\,(F_\al)_{jl}\over\tr F_\al}
=h(t_{kl}^\al\,(t_{ij}^\al)^*),
\cr}
$$
where we also used the chosen normalization for
$F_\al$.\hhalmos

\bPP
The last result justifies the normalization of the $F_\al$.
Note on the other hand that Propositions \thtag{38} and \thtag{39},
together with the property $f_1f_{-1}=\ep=f_{-1}f_1$ completely
determine $f_1$ and $f_{-1}$ as elements of $\FSA^*$.
The other $f_z$ are then determined by Proposition \thtag{37}(d),(e),
in view of Remark \thtag{36}.

A linear functional $f$ on an algebra $\FSA$ is called {\sl central} if
$f(ab)=f(ba)$ for all $a,b\in\FSA$.
Thus $h$ is not necessarily central, but its non-centrality is
well controlled by \eqtag{45}.

\Lemma{40}
For $a,b\in\FSA$ we have
$$
f_1.(ab).f_1=(f_1.a.f_1)\,(f_1.b.f_1).
\eqno\eq{49}
$$
\Proof
Apply \eqtag{45}:
$$
h(c\,(f_1.(ab).f_1))=h(abc)=h(bc\,(f_1.a.f_1))=
h(c\,(f_1.a.f_1)\,(f_1.b.f_1)).
$$
Now use that $d\in\FSA$ equals 0 if $h(cd)=0$ for all $c\in\FSA$.\hhalmos

\bPP
A linear functional $f\in\FSA^*$ is called {\sl multiplicative} if
$f(ab)=f(a)\,f(b)$ and $f(1)=1$. The multiplicative linear functionals
form again an algebra. (This is true for $\FSA$ an arbitrary bialgebra.)
Indeed, if $f,g\in\FSA$ are multiplicative then
$$
\displaylines{
\qquad(fg)(ab)=\sum_{(a),(b)}f(a_{(1)}b_{(1)})\,g(a_{(2)}b_{(2)})\hfill\cr
\hfill=
\sum_{(a),(b)}f(a_{(1)})\,g(a_{(2)})\,f(b_{(1)})\,g(b_{(2)})=
(fg)(a)\,(fg)(b).\qquad
\cr}
$$

\Prop{42}
For all $z\in\CC$ $f_z$ is a multiplicative linear functional:
$$
f_z(ab)=f_z(a)\,f_z(b),\quad a,b\in\FSA.
\eqno\eq{50}
$$
\Proof
Apply $\ep$ to both sides of \eqtag{49}.
Then we get $f_1^2(ab)=f_1^2(a)\,f_1^2(b)$.
Hence \eqtag{50} is true for $z=2$.
By the above remark, \eqtag{50} is then true for $z=2,4,6,\ldots\;$.
Now apply Proposition \thtag{37}(e) and Remark \thtag{36}.\hhalmos

\bPP
Proposition \thtag{42} tells us in particular that $f_1$ is
multiplicative. I wonder if a more algebraic proof, not using Carlson's
theorem, is possible for this fact.

\Remark{1000}
As a consequence of this Proposition, $f_1$ and $f_{-1}$ are
already determined in a CMQG algebra by their values on the matrix
elements $v_{ij}$, where $v$ is as in Proposition \thtag{28}(ii).
This also determines the \RHS s of \eqtag{47} and \eqtag{48},
since $(F_\al^{\pm1})_{ij}=f_{\pm1}(t_{ij}^\al)$.

\bPP
Let us finally consider expressions for $S(f_z)$ and $(f_z)^*$.
Note that a nonzero multiplicative linear functional $f\in\FSA^*$
has two-sided inverse $S(f)$. (This is true for $\FSA$ an arbitary Hopf
algebra.) Indeed,
$$
\displaylines{
\quad\ep(a)=f(\ep(a)1)=f(m\circ(S\ten\id)\De(a))\hfill\cr
\hfill=(f\ten
f)((S\ten\id)\De(a)) =(S(f)\ten f)(\De(a))=(S(f)\,f)(a).\quad
\cr}
$$
This shows that
$$
S(f_z)=f_{-z}.
\eqno\eq{51}
$$
In order to compute $(f_z)^*$ observe that
$$
(f_z)^*(t_{ij}^\al)
=\wb{f_z((S(t_{ij}^\al))^*)}
=\wb{f_z(t_{ji}^\al)}
=\wb{(F_\al^z)_{ji})}=(F_\al^{\wb z})_{ij}=f_{\wb z}(t_{ij}^\al).
$$
Hence
$$
(f_z)^*=f_{\wb z}.
\eqno\eq{59}
$$

\bLP
{\bf \the\sectionnumber.5. Comparison with other literature}
\sLP
(a) {\sl Woronowicz \ref{Wor87}, \ref{Wor88}, \ref{Wor91}}
\LP
Woronowicz, in his influential 1987 paper \ref{Wor87},
gives the following definition of a {\sl compact matrix quantum group}
(originally called {\sl compact matrix pseudogroup}).
It is a pair $(A,u)$, where $A$ is a unital $C^*$-algebra and
$u=(u_{ij})_{i,j=1,\ldots,N}$ is an $N\times N$ matrix with entries in $A$,
such that the following properties hold.
\item{1)}
The unital $*$-subalgebra $\FSA$ of $A$ generated by the entries of $u$
is dense in $A$.
\item{2)}
There exists a (necessarily unique)
$C^*$-homomorphism $\De\colon A\to A\ten A$ \st\
$\De(u_{ij})=\sum_{k=1}^N u_{ik}\ten u_{kj}$.
\item{3)}
There exists a (necessarily unique)
linear antimultiplicative mapping $S\colon \FSA\to\FSA$ \st\
$S\circ *\circ S\circ *=\id$ on $\FSA$ and
$\sum_{k=1}^N S(u_{ik})\,u_{kj}=\de_{ij}\,1=
\sum_{k=1}^N u_{ik}\,S(u_{kj})$.
\sLP
In his note \ref{Wor91} Woronowicz shows that, instead of 3),
we may equivalently require:
\item{3')}
The matrix $u$ and its transpose are invertible.
\mLP
Woronowicz now essentially shows (cf.\ \reff{Wor87}{Prop.\ 1.8})
that there exists a (necessarily unique)
$*$-homomorphism $\ep\colon\FSA\to\CC$ \st\ $\ep(u_{ij})=\de_{ij}$
and that $\FSA$ becomes a Hopf $*$-algebra with comultiplication $\De$,
counit $\ep$ and antipode $S$. In \ref{Wor87} the notation $\Phi,e,\ka$
is used instead of our $\De,\ep,S$, respectively.
Note that the above $*$-algebra $\FSA$ is very close to what
we have defined as
a CMQG algebra (cf.\ Proposition \thtag{26}). However, it is not postulated
and not yet obvious in the beginning of \ref{Wor87} that the co\rep s
$u$ and $u'$ are unitarizable.

A central result in the paper (see \reff{Wor87}{Theorem 4.2})
is the existence of a {\sl state}
(normalized positive linear functional) $h$ on the $C^*$-algebra $A$
\st\ $(h\ten\id)(a)=h(a)\,1=(\id\ten h)(a)$ for all $a\in A$.
This state is necessarily unique and it is faithful on $\FSA$.
Then $h$ may be called the Haar functional.

Woronowicz \reff{Wor87}{\S2}
defines a {\sl representation} of the
compact matrix quantum group $(A,u)$ on a finite dimensional vector
space $V$
as a linear mapping
$t\colon V\to V\ten A$ \st\ $(t\ten\id)\circ t=(\id\ten\De)\circ t$.
If $t(v)=0$ implies $v=0$ then the \rep\ is called {\sl non-degenerate}
and if $t(V)\i V\ten\FSA$ then the \rep\ is called {\sl smooth}.
A smooth \rep\ is non-degenerate iff $(\id\ten\ep)\circ t=\id$.
Thus co\rep s of $\FSA$ on finite dimensional vector spaces, as defined
in \S1.3 of the present paper, correspond to nondegenerate smooth
\rep s of $(A,u)$ in \ref{Wor87}.

As a consequence of the existence of the Haar functional, it is now shown
(see \reff{Wor87}{Theorem 5.2 and Prop.\ 3.2}) that nondegenerate smooth
\rep s of $(A,u)$ are unitarizable. Then we can conclude that the
dense $*$-algebra $\FSA$ of $A$ is a CMQG algebra.

Conversely, if we start with a CMQG algebra $\FSA$ with fundamental co\rep\
$u$ as in Proposition \thtag{26}
then it is possible
to show the existence of a Haar functional $h$
(cf.\ Theorem \thtag{33}) without using $C^*$-algebras.
Next a $C^*$-completion $A$ can be made which uses the existence of the Haar
functional (cf.\ \S\the\sectionnumber.3).
Then it is clear that the pair $(A,u)$ is a compact matrix quantum group
in the sense of Woronowicz.
However, the $C^*$-algebra $A$ as constructed from $\FSA$ in
\S\the\sectionnumber.3, is canonically determined by $\FSA$, but this is not
the case with the compact matrix quantum groups $(A,u)$ of Woronowicz.
In the present paper we have taken the completion of $\FSA$ \wrt the
largest $C^*$-seminorm on $\FSA$, while, in \ref{Wor87}, the norm induced
by $A$ on $\FSA$ may not be the largest $C^*$-seminorm on $\FSA$.
In that case, the counit $\ep\colon \FSA\to\CC$ does not necessarily have
a continuous extension to a linear functional on $A$
(cf.\ \reff{Wor87}{second Remark to Prop.\ 1.8}).

In his paper \ref{Wor88}, Woronowicz starts with a compact matrix
quantum group
$(A,u)$ in the sense of \ref{Wor87}, then constructs out of its finite
dimensional unitary representations a so-called complete concrete
monoidal $W^*$-category (see \reff{Wor88}{Theorem 1.2}) and next
constructs from any such category a compact matrix quantum group $(B,u)$.
Then $B$ is not necessarily isomorphic to $A$, but $B$ is obtained
canonically from the CMQG algebra $\FSA$ generated
by the entries of $u$, as in
\S\the\sectionnumber.3.

Both in \ref{Wor87} and in the present paper there is a similar key result
\reff{Wor87}{last statement of Theorem 5.4} resp.\
Proposition \thtag{32}. We got the idea of the statement and proof
of Proposition \thtag{32} from \ref{Wor87}, but in the present paper,
different from \ref{Wor87},
the positivity and faithfulness of the Haar functional
on $\FSA$ is a corollary
rather than a prerequisite.
As soon as these matters are settled, the quantum Schur orthogonality
relations and the modular property of the Haar functional can be obtained
in the same way in both papers, cf.\ \reff{Wor87}{pp.\ 653--656}
resp.\ \S\the\sectionnumber.4 in the present paper.

\mLP
(b) {\sl Woronowicz  \ref{Wor93} and S. Wang \ref{Wan93}}
\LP
Woronowicz \ref{Wor93} defines a {\sl compact quantum group}
as a pair $(A,\De)$, where $A$ is a separable unital $C^*$-algebra and
$\De\colon A\to A\ten A$ is a $C^*$-homomorphism, such that the following
properties hold.
\item{1)}
$(\De\ten\id)\circ\De=(\id\ten\De)\circ\De$.
\item{2)}
$\Span\{(b\ten 1)\,\De(c)\mid b,c\in A\}$
and $\Span\{(1\ten b)\,\De(c)\mid b,c\in A\}$
are dense subspaces of $A\ten A$.

\sLP
In particular, if $(A,u)$ is a compact matrix pseudogroup as defined in
\ref{Wor87} and if $\De$ is the corresponding comultiplication
then $(A,\De)$ is a compact quantum group as just defined.
Conversely, it is shown in \ref{Wor93} that, if $(A,\De)$ is a compact
quantum group and if $\FSA$ is the set of all linear combinations
of matrix elements of finite dimensional unitary
representations of $(A,\De)$
then $\FSA$ is a dense $*$-subalgebra
of $A$ and $\FSA$ is a Hopf $*$-algebra.
The existence of a Haar functional is also shown.
It is observed that the representation theory as developed in \ref{Wor87}
can be formulated in a similar way for compact quantum groups.

It is pointed out in Wang \reff{Wan93}{Remark 2.2} that the results of
\ref{Wor93} remain true if separability of the
$C^*$-algebra $A$ is no longer
required, but if it is assumed instead that the $C^*$-algebra $A$
has a faithful state.
This observation would imply that a compact quantum
group $(A,\De)$ in the
sense of Wang gives rise to a CQG algebra $\FSA$ ($\FSA$ being
constructed from $A$ as in the previous paragraph),
and that conversely each
CQG algebra $\FSA$ would give rise to a compact quantum group $(A,\De)$
($A$ being completion of $\FSA$ \wrt maximal $C^*$-seminorm).

\mPP\goodbreak\noindent
(c) {\sl Effros \& Ruan \ref{EfRu93}}
\nobreak\LP
In different terminology,
CQG algebras were earlier introduced by Effros \& Ruan \ref{EfRu93}.
They defined these algebras as cosemisimple Hopf algebras with
a so-called standard $*$-operation and
they called these structures {\sl discrete quantum groups}.
This name was motivated by the fact that special examples of these algebras
are provided by the group algebra of a discrete group,
while the name CQG algebra comes from the class of examples, where we
deal with the algebra
of representative functions on a compact group.
In the final section of \ref{EfRu93} the authors define a
{\sl compact quantum group} as a natural generalization of the compact
matrix quantum groups defined in \ref{Wor87}. Their definition involves
a unital $C^*$-algebra $A$ with a dense unital $*$-subalgebra $\FSA$,
where $\FSA$ is a CQG-algebra (in the terminology of the present paper)
and the comultiplication on $\FSA$ extends continuously to $A$.
Conversely, they show that a CQG algebra $\FSA$ gives rise to a
compact matrix
quantum group according to their definition.
This involves a $C^*$-completion, for which a Haar functional $h$ on $\FSA$
is needed. This Haar functional is obtained in a way very different from
the method in the present paper. The authors first show the existence
of a left Haar functional $\phi$ on a certain subspace
of the linear dual of $\FSA$. Then $h$ is constructed in terms of $\phi$.

\Ref
\refitem{Abe}
E. Abe, {\sl Hopf algebras}, Cambridge University Press, 1980.

\refitem{Arv76}
W. Arveson,
{\sl An invitation to $C^*$-algebra},
Springer, 1976.

\refitem{Dij94}
M. S. Dijkhuizen,
{\sl On compact quantum groups and quantum homogeneous spaces},
Dissertation, University of Amsterdam, 1994.

\refitem{EfRu93}
E. G. Effros \&
Z.-J. Ruan,
{\sl Discrete quantum groups, I. The Haar measure},
preprint, 1993.

\refitem{Haz78}
M. Hazewinkel,
{\sl Formal groups and applications},
Academic Press, 1978.

\refitem{Swe}
M. E. Sweedler, {\sl Hopf algebras}, Benjamin, 1969.

\refitem{Tit39}
E. C. Titchmarsh,
{\sl The theory of functions},
Oxford University Press, second ed., 1939.

\refitem{Wan93}
S. Z. Wang,
{\sl General constructions of compact quantum groups},
preprint, 1993.

\refitem{Wor87}
S. L. Woronowicz,
{\sl Compact matrix pseudogroups},
Comm. Math. Phys. 111 (1987), 613--685.

\refitem{Wor88}
S. L. Woronowicz,
{\sl Tannaka-Krein duality for compact matrix pseudogroups.
Twisted SU(N) groups},
Invent. Math. 93 (1988), 35--76.

\refitem{Wor91}
S. L. Woronowicz,
{\sl A remark on compact matrix quantum groups},
Lett. Math. Phys. 21 (1991), 35--39.

\refitem{Wor93}
S. L. Woronowicz,
{\sl Compact quantum groups},
preprint, version August 1, 1993.

\par
\bye